\documentclass[twoside,12pt]{article}
\usepackage{epsfig}
\usepackage{color}
\usepackage{hyperref}

\newcommand{\be}{\begin{equation}}
\newcommand{\ee}{\end{equation}}
\newcommand{\bea}{\begin{eqnarray}}
\newcommand{\eea}{\end{eqnarray}}

\begin{document}
%%%%%%%%%%%%%%%%%%%%%%%%%%%%%%%%%%%%%%%%%%%%%%%%%%%%%%%%%%%%%%%%%
\title{ \vspace{1cm}  Current unknowns in the three-neutrino framework}
\author{F.~Capozzi,$^{1}$
E.~Lisi,$^{2}$
A.~Marrone,$^{3,2}$
A.~Palazzo,$^{3,2}$\medskip\\
\small\em
$^1$Max-Planck-Institut f\"ur Physik (Werner-Heisenberg-Institut), F\"ohringer Ring 6, 80805 M\"unchen, Germany
\\
\small\em $^2$ Istituto Nazionale di Fisica Nucleare, Sezione di Bari, Via Orabona 4, 70126 Bari, Italy 
\\
\small\em $^3$Dipartimento Interateneo di Fisica dell'Universit\`a di Bari, Via Amendola 173, 70126 Bari, Italy
}
\maketitle
\begin{abstract}  
We present an up-to-date global analysis of data coming from neutrino oscillation and non-oscillation experiments, 
as available in April 2018, within the standard framework including three massive and mixed neutrinos. We discuss in
detail the status of the three-neutrino ($3\nu)$ mass-mixing parameters, both known and unknown. 
Concerning the latter, we find that: normal ordering (NO) is favored over inverted ordering (IO) at $3\sigma$ level;  
the Dirac CP phase is constrained within $\sim 15\%$ ($\sim 9\%$) uncertainty in NO (IO) around nearly-maximal CP-violating values;
the octant of the largest mixing angle and the absolute neutrino masses remain undetermined. We briefly comment on other  unknowns related to
theoretical and experimental uncertainties (within $3\nu)$  
or possible new states and interactions (beyond $3\nu$). 
\end{abstract}
%\eject
%\tableofcontents

\section{Introduction}

This work represents an ideal follow-up of a previous review in this Journal \cite{Fogli:2005cq},
where a global analysis of oscillation and non-oscillation data as of 2005 was discussed in detail, within the 
framework of three massive and mixed neutrinos ($3\nu$). This framework, that has gradually emerged from a series of beautiful 
experiments, 
%performed with neutrinos from both natural sources (atmospheric and solar) and man-made ones 
%(reactor and accelerator), 
represents now a ``standard'' paradigm of particle physics 
\cite{Patrignani:2016xqp,Petcov}, as also highlighted by the Nobel Prize in Physics 2015 \cite{Nobel2015},
that crowned decisive oscillation discoveries with natural (atmospheric and solar) neutrinos
\cite{Kajita:2016cak,McDonald:2016ixn}, and by the Breakthrough Prize in Fundamental 
Physics 2016 \cite{Breakthrough2016}, 
awarded to milestone experiments using both natural and man-made (reactor and accelerator) neutrino beams
\cite{Cao:2017drk,Diwan:2016gmz}.

The three-neutrino paradigm is based on the simplest assumption beyond massless neutrinos, namely,   
that the three known flavor states $\nu_\alpha=(\nu_e,\,\nu_\mu,\,\nu_\tau)$ are linear combinations of three
states $\nu_i=(\nu_1,\,\nu_2,\,\nu_3)$ with definite masses $m_i=(m_1,\,m_2,\,m_3)$ through a unitary matrix $U_{\alpha_i}$,
also called the Pontecorvo-Maki-Nakagawa-Sakata \cite{Pontecorvo:1967fh,Maki:1962mu}  (PMNS)  matrix \cite{Giganti:2017fhf}. 
In standard convention \cite{Petcov}, $U_{\alpha_i}$ is 
parameterized in terms of three mixing angles $\theta_{ij}\in [0,\pi/2)$ and one so-called Dirac phase $\delta\in [0,\,2\pi)$, associated to
possible violations of the charge-parity  (CP) symmetry in the neutrino sector,
%-------------------------------
\begin{equation}
U_{\alpha i}=\left(
\begin{array}{ccc}
c_{13}c_{12} & s_{12}c_{13} & s_{13}e^{-i\delta}\\
-s_{12}c_{23}-c_{12}s_{23}s_{13}e^{i\delta} &  c_{12}c_{23}-s_{12}s_{23}s_{13}e^{i\delta} & s_{23}c_{13}\\
 s_{12}s_{23}-c_{12}c_{23}s_{13}e^{i\delta} & -c_{12}s_{23}-s_{12}c_{23}s_{13}e^{i\delta} & c_{23}c_{13}\\
\end{array}
\right)\ ,
\label{PMNS}
\end{equation}
%-------------------------------
where $c_{ij}=\cos(\theta_{ij})$ and $s_{ij}=\sin(\theta_{ij})$.

A convention-independent measure of CP violation is given by the Jarlskog invariant \cite{Jarlskog:1985ht,Petcov}
%----------------------------
\begin{eqnarray}
J &=&\mathrm{Im}\left( U_{\mu3}U^{\tt *}_{e3}U_{e2}U_{\mu 2}^{\tt *}\right)\\
&=& \frac{1}{8}\cos(\theta_{13})\sin(2\theta_{13})\sin(2\theta_{23})\sin(2\theta_{12})\sin\delta\ ,
\end{eqnarray}
%--------------------------------   
which shows at a glance that leptonic CP violation is a genuine $3\nu$ effect \cite{Cabibbo:1977nk}. In particular, $J\neq 0$ requires not only that 
$\delta \neq \{0,\,\pi\}$ but also that any mixing angle is nonzero ($\theta_{ij}> 0$) and that any two masses 
are different ($m_i\neq m_j$) -- otherwise one mixing angle could be rotated away. Since the review
in \cite{Fogli:2005cq}  (when only $\theta_{12}$ and $\theta_{23}$ were measured), dramatic progress has occurred 
concerning $\theta_{13}$ \cite{Apollonio:2002gd}, 
starting from hints from solar, reactor and atmospheric data 
\cite{Fogli:2008jx}, to growing evidence from 
accelerator data \cite{Abe:2011sj,Adamson:2011qu,Fogli:2011qn,Schwetz:2011zk}
 and finally to {its} discovery and precise determination via near-far detection at short-baseline reactors 
 \cite{Abe:2011fz,An:2012eh,Ahn:2012nd,Tortola:2012te,Fogli:2012ua,GonzalezGarcia:2012sz}.  
Currently, not only the $J$ prefactor is known to be nonzero, but very interesting data from 
long-baseline accelerator experiments \cite{Abe:2013hdq,Adamson:2016tbq,Abe:2017uxa,Adamson:2017gxd}
 seem to suggest a nearly maximal CP phase factor, $|\sin\delta|\sim 1$, with
a significant preference for $\sin\delta <0$ \cite{Esteban:2016qun,nufit,deSalas:2017kay,Capozzi:2016rtj,Capozzi:2017ipn}. 
Another issue  is the near maximality of the $\theta_{23}$ angle, $\sin^2(2\theta_{23})\simeq 1$, that is still unresolved
\cite{Abe:2013hdq,Adamson:2016tbq,Abe:2017uxa,Adamson:2017gxd}
 in terms of a preferred octant ($\theta_{23}\leq \pi/4$ or $> \pi/4$) 
\cite{Fogli:1996pv}.

Concerning neutrino mass states $\nu_i$, the standard labelling \cite{Petcov} 
$i=1,\,2,\,3$ reflects the observed hierarchy of $\nu_e$ mixing with $\nu_i$, namely, $|U_{e1}|^2>|U_{e2}|^2>|U_{e3}|^2$. In vacuum, 
oscillations of relativistic $\nu_i$ with given momentum $p$ and energies $E(\nu_i)\simeq p+m^2_i/(2p)$ are triggered 
by the tiny differences $E(\nu_i)-E(\nu_j)\simeq (m^2_i-m^2_j)/(2p)$. Using the same convention as in \cite{Fogli:2005cq} we define 
two independent squared mass differences,
%----------------------
\begin{equation}
\delta m^2 = m^2_2-m^2_1 > 0
\end{equation}
%-------------------------
and
%------------------------
\begin{equation}
\Delta m^2 = m^2_3 -\frac{m^2_1+m^2_2}{2}\ ,
\end{equation}
%----------------------------
with two possible options for the neutrino mass spectrum ordering: either $\Delta m^2>0$ (normal ordering, NO) or $\Delta m^2<0$ (inverted ordering, IO).
In matter, oscillations of $\nu_\alpha$ are also affected by interaction energy 
differences $E(\nu_e)-E(\nu_{\mu,\tau})=\sqrt{2}G_FN_e$ via
the celebrated Mikheev-Smirnov-Wolfenstein (MSW) mechanism \cite{Wolfenstein:1977ue,Mikheev:1986gs,Mikheev:1986wj} 
and its variants for different profiles of the electron density $N_e(x)$ \cite{Petcov,Blennow:2013rca}. 
Oscillations in both vacuum and matter have led to measurements of $\delta m^2$ and $|\Delta m^2|$ but not yet of the sign of $\Delta m^2$
\cite{Patterson:2015xja}.  

Finally, absolute neutrino masses are accessible via kinematical effects at the endpoint of $\beta$ decay
\cite{Otten:2008zz,Drexlin:2013lha}, approximately sensitive to 
the so-called effective electron neutrino mass $m_\beta$ defined as \cite{Mbet},
%-------------------
\begin{equation}
m^2_{\beta} =  \sum_{i=1}^3 |U_{ei}|^2 m^2_i = c^2_{13}(c^2_{12}m^2_1+s^2_{12}m^2_2)+s^2_{13}m^2_3 \ ,
\end{equation}
%-----------------------
or via dynamical effects from gravitational and electroweak interactions. 
In particular, cosmological observations are sensitive to the sum of neutrino masses (i.e., to their ``total gravitational charge'') 
\cite{Dolgov:2002wy,Hannestad:2010kz,Wong:2011ip,Lesgourgues:2012uu},
%-----------------------
\begin{equation}
\Sigma = m_1+m_2+m_3\ ,
\end{equation}
%------------------------
and, to some extent, also to the mass spectrum ordering \cite{Lattanzi:2017ubx}.
If neutrinos are of Majorana (instead of Dirac) type \cite{Majorana:1937vz,Petcov:2013poa}, 
then a rare process of two-lepton creation---the neutrinoless double beta decay
($0\nu\beta\beta$)---may occur in some nuclei \cite{Bilenky:2014uka,Pas:2015eia,DellOro:2016tmg,Vergados:2016hso}, 
with a rate proportional to the square of the effective Majorana mass $m_{\beta\beta}$,
%--------------------------
\begin{equation}
m_{\beta\beta}=\left| \sum_{i=1}^3 U^2_{ei} m_i\right| = \left| c^2_{13}(c^2_{12}m_1+s^2_{12}e^{i\phi_{21}}m_2)+s^2_{13}e^{i\phi_{31}}m_3 \right|\ ,
\end{equation}
%-------------------------
where $\phi_{ji}$ are additional (Majorana) CP-violating phases, here defined via the convention 
$U \to U \cdot \mathrm{diag}(1,\,e^{\frac{i}{2}\phi_{21}},\,e^{\frac{i}{2}(\phi_{31}+2\delta)})$ \cite{Vogel}. 
See also \cite{Fantini:2018itu} for an interesting recent overview of  
formalism and conventions in neutrino physics.

Within the above $3\nu$ framework of massive and mixed neutrinos, we currently know rather accurately five parameters, governing two
oscillation frequencies and their amplitudes in different channels, 
%-----------------
\begin{equation}
\label{eq:knowns}
3\nu\ \mathrm{knowns}:  \ \delta m^2,\,|\Delta m^2|,\,\theta_{12},\,\theta_{23},\,\theta_{13}\ , 
\end{equation}
%--------------------
while the following five features have not been established yet:
%---------------------
\begin{equation}
\label{eq:unknowns}
3\nu\ \mathrm{unknowns}:\  \delta,\,\, \mathrm{sign}(\Delta m^2),\,\mathrm{sign(\theta_{23}-\pi/4)},\, \mathrm{min}(m_i),\,
 \mathrm{Dirac/Majorana\ nature}\ ,
\end{equation}
%-----------------------
the latter option including the unknown phases $\phi_{21}$ and $\phi_{31}$ (if Majorana). In the following we shall review the status of both 
known and unknown features of the $3\nu$ framework, within a global analysis of oscillation and non-oscillation data as available in April 2018. 
Results will be expressed in terms of standard 
deviations $N\sigma$  from a local or global $\chi^2$ minimum, 
%---------------------
\begin{equation}
N{\sigma} = \sqrt{\Delta \chi^2}\ .
\end{equation}
%-----------------------
This analysis follows up the previous review in this Journal \cite{Fogli:2005cq} and updates the more recent papers in 
\cite{Capozzi:2016rtj,Capozzi:2017ipn}. 
Interesting and independent global analyses of neutrino data have 
also appeared recently \cite{Esteban:2016qun,nufit,deSalas:2017kay}, and will be referred to for comparison in the following.

There are also other ``unknowns'' (or poorly known quantities) that affect  the completion and test of
the $3\nu$ framework. On the one hand, several physics ingredients demand a deeper experimental and theoretical knowledge,
at the level of neutrino production (e.g., absolute fluxes and energy spectra), evolution in time 
(e.g., background fermion profiles in matter, large-scale structure effects in cosmology), and detection (e.g., 
absolute and differential cross sections, effective weak couplings in nuclear matter). On the other hand, at any given
time there are some observed phenomena that seem to go beyond the adopted ``standard neutrino framework,'' and that might point towards novel states or interactions. A relevant example is currently provided by anomalous oscillation results suggesting mixing with 
light sterile neutrinos \cite{Palazzo:2013me,Gariazzo:2015rra,Gariazzo:2017fdh,Dentler:2018sju}, 
possibly endowed with peculiar interactions to evade cosmological bounds \cite{Hannestad:2012ky,Forastieri:2017oma,Archidiacono:2016kkh,Chu:2015ipa}.     
An overview of these ``generalized'' unknown (or poorly known) aspects of neutrino physics  
is beyond the scope of this paper, but we shall briefly comment on some
of them while discussing the ``proper'' $3\nu$ knowns and unknowns in Eqs.~(\ref{eq:knowns}) and (\ref{eq:unknowns}),
especially to highlight $3\nu$ aspects which deserve further attention. 

The paper is organized as follows. In Section~2 we discuss the methodology and the updates used to analyze solar,
long-baseline reactor, long-baseline accelerator, short-baseline reactor, and atmospheric $\nu$ oscillation data.
With respect to \cite{Fogli:2005cq} and also to more recent analyses \cite{Capozzi:2016rtj,Capozzi:2017ipn} we now use official 
$\chi^2$ maps provided by experimental collaborations for some data sets (including 
short-baseline reactor and atmospheric neutrinos) that
are difficult to reproduce by external users, and discuss pros and cons of 
this choice.
In Section~3 we discuss the global fit results in terms of single (known and unknown) oscillation parameters. We find
persisting hints in favor of $\sin\delta<0$ and significant indications in favor of normal spectrum ordering, at the level of $N\sigma\simeq 3$. 
Concerning $\theta_{23}$, we find a rather
restricted range near maximal mixing, with a slight preference for the second octant.
In Section~4 we explore in further detail such results in terms of covariances between pairs of parameters, which highlight 
the interplay among different data sets. In Section 5 we combine oscillation data 
with nonoscillation constraints from cosmology and neutrinoless double beta decay, 
in order to derive upper bounds on absolute neutrino masses, which are of interest also for neutrino mass searches with beta decay. 
Our summary and conclusions are reported in Section 6.

\newpage
A final remark is in order. As in \cite{Fogli:2005cq}, we aim at presenting a state-of-the-art  global analysis of $3\nu$ knowns
and unknowns, but we do not aim at being bibliographically complete. A useful starting point for orientation in the vast neutrino literature
is \cite{Unbound}.
Recent books with useful references on various aspects of
our current understanding (and future prospects) of the neutrino mass-mixing phenomenology and its relation with astroparticle physics and cosmology
include \cite{Giunti2007,Barger2012,Miele2013,Suekane2015,Spurio2015,Valle2015,Ohlsson2016,Sigl2017,Aloisio2018,Ereditato2018,Bilenky2018}. With
respect to \cite{Fogli:2005cq}, we do not insist anymore on statistical aspects that have now become standard tools in the field (such 
as details of the pull method \cite{Fogli:2002pt} and of its various applications in $\chi^2$ analyses), but prefer to comment on future challenges
that are emerging from  analyses of current and prospective data. 

%%%%%%%%%%%%%%%%%%%%%%%%%%%%%%%%%%%%%%%%%%%%%%%%%%%%%%%%%%%%%%%%%%%%%%%%%%%%%%%%%%%%%%%%%%%%%%%%%%%%%%%%%
\section{Analysis of oscillation data: Methodology and updates}

In the review \cite{Fogli:2005cq}, it was found that the mixing angle $\theta_{13}$ was compatible with zero at $\sim 1\sigma$, 
although its best-fit value ($\sin^2\theta_{13} \sim 10^{-2}$) was already in the right ballpark of later discoveries. 
At that time, under the assumption $\theta_{13}\simeq 0$ suggested by the CHOOZ null results \cite{Apollonio:2002gd}, it
was methodologically convenient---before performing a global fit---to group oscillation data in two classes: solar plus long-baseline reactor data, mainly  
sensitive to $(\delta m^2,\,\theta_{12})$, and atmospheric and long-baseline accelerator data, mainly sensitive to $(\Delta m^2,\,\theta_{23})$ 
\cite{Fogli:2005cq}.

Subsequently, the growing indications in favor of $\sin^2\theta_{13}\sim 0.01$--0.02 \cite{Fogli:2008jx}
and its experimental discoveries via $\nu_e\to\nu_e$ flavor disappearance at reactors \cite{Cao:2017drk}
and $\nu_\mu\to\nu_e$ appearance at accelerators \cite{Diwan:2016gmz},
have opened a portal to genuine $3\nu$ effects at subleading level \cite{Fogli:2012ua,Capozzi:2013csa}, 
such as possible signs of CP violation driven by $\delta$ \cite{Abe:2017uxa}
 and, more recently, to possible indications about the mass ordering \cite{Abe:2017aap}; 
 see also \cite{Esteban:2016qun,nufit,deSalas:2017kay,Capozzi:2016rtj,Capozzi:2017ipn}. 
 In this context, a different grouping of data was proposed in \cite{Fogli:2012ua},
in order to show more clearly the progressive impact of different data sets on both known and unknown parameters. The same methodology is also 
adopted herein, being supported by additional reasons that we now discuss.

The starting point is provided by solar and long-baseline reactor data, that probe the $\nu_e\to\nu_e$ flavor disappearance channel 
via oscillations driven by the $(\delta m^2,\theta_{12},\theta_{13})$ parameters. These data provide precise measurements  
of $(\delta m^2,\theta_{12})$ and a rough measurement of $\theta_{13}>0$ at the $\sim2\sigma$ level. On the other hand, long-baseline accelerator
data probe both the $\nu_\mu\to\nu_\mu$ disappearance and the $\nu_\mu\to\nu_e$ appearance channel via oscillations driven mainly by
the $(\Delta m^2,\,\theta_{23},\,\theta_{13})$ parameters, but they are also sensitive to subleading effects driven by $(\delta m^2,\theta_{12})$, 
as well as by $\delta$ and $\mathrm{sign}(\Delta m^2)$. The interesting new fact is that such  data provide, in combination, not
only a good measurement for each of the five parameters in Eq.~(\ref{eq:knowns}), but also precious hints in favor of $\sin\delta \neq 0$ 
(i.e., CP violation) and of $\mathrm{sign}(\Delta m^2)=+1$ (i.e., normal ordering), as shown in Sec.~3.  
It turns out that these hints are enhanced by adding first short-baseline reactor data, mainly sensitive to $(\Delta m^2,\,\theta_{13})$,
and then atmospheric neutrino data, sensitive in different ways to all the oscillation parameters via disappearance and appearance
channels.  Therefore, this methodological approach 
allows to gauge how the current indications about neutrino CP violation and mass spectrum ordering are progressively enhanced, 
by using increasingly rich data sets in the global analysis.

In the following, we discuss relevant updates for the various data sets, in the same order as suggested by the above methodology.  
The reader not interested in technical details may skip the rest of this Section and  jump to the results in Section~3.

\subsection{Solar and long-baseline reactor (KamLAND) neutrinos}

Concerning solar neutrinos, with respect to the recent analyses in \cite{Capozzi:2016rtj,Capozzi:2017ipn} 
we now include the latest low-energy Borexino data \cite{Bellini:2011rx,Bellini:2014uqa,Agostini:2017ixy} and
Super-Kamiokande-IV data (day and night) \cite{Abe:2016nxk}. 
We have revisited our analysis of three-phase data from the Sudbury Neutrino Observatory (SNO), 
obtaining results in closer agreement with the official SNO ones reported in \cite{Aharmim:2011vm} 
for the mass-mixing parameter region allowed
at large mixing angle (LMA). 
Inputs from radiochemical experiments remain as reported in \cite{Cleveland:1998nv} (Homestake) for Chlorine 
and in \cite{Abdurashitov:2009tn,Kaether:2010ag} (GALLEX-GNO + SAGE) for Gallium.

We adopt reference solar neutrino fluxes from the standard solar model named B16-GS98  in 
\cite{Vinyoles:2016djt}, which ameliorates the tension
 with helioseismological data. Systematic nuisance uncertainties are taken from \cite{Vinyoles:2016djt} 
 and, when needed, they are
supplemented by related information from \cite{Serenelli:2012zw}. 
The $^{8}$B neutrino spectrum and its uncertainties are taken from \cite{Winter:2004kf}.

The Gallium (Ga) neutrino absorption cross-section $\sigma_\mathrm{Ga}(E_\nu)$ and its uncertainties have been updated according to
the recent experimental results and estimates in \cite{Frekers} (see also \cite{Barinov:2017ymq}).   
The results of \cite{Frekers} lead to a reduction of the unoscillated solar neutrino rate in Ga by $\sim 6$~SNU (solar neutrino units \cite{Bahcall}) 
for our adopted reference fluxes \cite{Vinyoles:2016djt} (the reduction by $\sim 10$~SNU quoted in \cite{Frekers} being due to somewhat different fluxes).  
The generic impact of Ga cross-section variations was discussed in \cite{Fogli:2006fu}, where it was shown
that, in combination with SNO, a reduction of $\sigma_\mathrm{Ga}$ tends to slightly decrease 
$\theta_{12}$ and increase $\delta m^2$ for nonzero $\theta_{13}$ (see Figs.~8 and 9 in \cite{Fogli:2006fu}). These qualitative expectations are
confirmed in our analysis.

Concerning long-baseline reactor neutrino oscillations in the KamLAND (KL) experiment \cite{KL13}, we 
adopt the same reanalysis of the 2011 KL data set \cite{KL11} performed
in \cite{Capozzi:2016rtj}, which included in 
the reactor spectra \cite{Huber:2011wv,Mueller:2011nm} the ``bump'' feature recently observed around energies 
$E_\nu\sim 5$--7~MeV
\cite{Seo:2014xei,RENO:2015ksa,Abe:2014bwa,An:2015nua}, which
is still poorly understood  \cite{Novella:2015eaw,Sonzogni:2015aoa,Hayes:2015yka,Huber:2016fkt}). Such a 
reanalysis has led to a tiny reduction of the $(\delta m^2,\,\theta_{12})$ 
best-fit values in KL \cite{Capozzi:2016rtj}, see also \cite{Maltoni:2015kca}. In this context, an official KL analysis update (including current information
on reactor spectra and uncertainties) would be beneficial.

As in \cite{Capozzi:2016rtj}, we cannot use the latest published KL data \cite{KL13} herein. They are presented in a peculiar format 
(consisting of three subsets with correlated systematics) that prevents a proper detailed
analysis outside the collaboration. This drawback is representative
of a more general situation that is becoming increasingly common in neutrino physics---as also discussed later in this review---and 
that parallels other fields of particle 
physics involving multiple and complex experimental inputs, such as global analyses of electroweak data \cite{Baak:2014ora,Alioli:2016fum}
and of parton distribution functions \cite{Alioli:2016fum,Dulat:2015mca,Harland-Lang:2014zoa}.  Indeed, as the experiments
become more refined and collect higher statistics with trickier dependence on systematics, the data analysis also gets more complicated, 
eventually becoming nearly prohibitive outside the collaborations. However, external users 
may need to perform their own 
data fits for diverse purposes, e.g., for global analyses as in this work, or for phenomenological tests of specific theoretical models,  
or for sensitivity estimates of prospective signals. One can then adopt different approaches to such situation,
with various pros and cons, including:
(a)  continue to analyze (some) available data within reasonable approximations or well-defined restrictions, but with increasing awareness of the 
inherent uncertainties; (b) discard raw data in favor of officially ``processed'' results (e.g., via $\chi^2$ maps or dedicated  
software tools, if any) that, however, may prevent 
testing analysis details or variants; (c) just give up on some data (sub)sets. Given the complex issues involved, 
one should maintain
an open attitude about different choices (that may be dictated by objective difficulties as well as by subjective assessments), 
and foster a continuous dialogue 
between internal collaboration teams and external researchers, so as to use the precious experimental data in the best possible way 
to advance neutrino phenomenology and theory. 

We conclude by reminding that the $3\nu$ survival probability relevant for solar and KamLAND neutrino data can be cast in
the form \cite{Petcov,Maltoni:2015kca}:
%------------
\begin{equation}
P_{3\nu}(\nu_e\to\nu_e) \simeq \cos^4\theta_{13}\,P_{2\nu}(\nu_e\to\nu_e) +\sin^4\theta_{13}\ ,
\label{P3nu2nu}
\end{equation}
%-------------- 
where $P_{2\nu}$ corresponds to the $2\nu$ probability for $\theta_{13}=0$, which depends on the $(\delta m^2,\,\theta_{12})$ parameters only.
For solar neutrinos, one should replace $\theta_{13}$ with its effective value in matter $\tilde \theta_{13}$, which carries a slight
dependence on $\Delta m^2$ and on  the mass ordering \cite{Fogli:2001wi,Fogli:2005cq}.

\subsection{Long-baseline accelerator neutrinos}

At the time of the previous review  \cite{Fogli:2005cq} on this Journal, long-baseline (LBL) accelerator searches for $\nu_\mu\to\nu_\mu$ 
disappearance had been performed only by 
the KEK-to-Kamioka (K2K) experiment \cite{Ahn:2002up,Aliu:2004sq}, 
later followed by the Main Injector Neutrino Oscillation Search (MINOS) experiment \cite{Adamson:2008qj} 
that also started to search for $\nu_\mu\to\nu_e$ appearance \cite{Adamson:2014vgd}. The successful search for 
for $\nu_\mu\to\nu_\tau$ appearance in the Oscillation Project with Emulsion-tRacking Apparatus (OPERA) \cite{Agafonova:2015jxn} has provided further confidence in the $3\nu$ framework, although the data in this channel do not  significantly constrain  the $3\nu$ oscillation parameters.

Two main experiments currently drive the search for both $\nu_\mu\to\nu_\mu$ and $\nu_\mu\to\nu_e$ oscillations with 
LBL accelerator neutrino and antineutrino  beams, namely, the Tokai-to-Kamioka (T2K) experiment
in Japan \cite{Abe:2017uxa,Abe:2017vif} and the Neutrino at main injector Off-axis $\nu_e$ Appearance 
experiment in the U.S.~\cite{Adamson:2017gxd}. A powerful software, the General Long Baseline Experiment Simulator (GLoBES) 
\cite{Huber:2004ka,Huber:2007ji}
has also become publicly available to analyze this class of experiments, including a full-fledged treatment of
statistical and systematic uncertainties. 
Previous analyses that included T2K and NOvA results via adapted versions of GLoBES have been discussed in 
\cite{Capozzi:2016rtj,Capozzi:2017ipn}.
 
Very recently, updated disappearance and appearance data have been presented for both T2K \cite{t2k_2017} and NOvA \cite{nova_2018}.   
For T2K \cite{t2k_2017}, disappearance data include 240 $\nu_\mu$ events and 68 $\overline\nu_\mu$ events in the 
charged-current quasi-elastic (CCQE) class, divided into 27 equally-spaced bins in 
the reconstructed energy  range $[0.2,\,2.9]$~GeV, plus a 28th bin (4 GeV wide) for higher energies. 
Appearance data include 74 CCQE $\nu_e$, 7 CCQE 
$\overline\nu_e$ and 15 CC$1\pi$ (one pion) $e$-like events, divided into 9 equally-spaced bins in the interval 
$[0.125,\,1.25]$~GeV. Background events are taken from \cite{t2k_2017} and assumed to be oscillation-independent.
In our analysis, software-generated disappearance and appearance spectra are calibrated so as to reasonably reproduce 
the official spectra at the oscillation best-fit point, which require $\sim 15\%$ energy resolution
smearing. Agreement with official parameter bounds is also optimized by slightly tuning nuisance parameters, 
including the normalization uncertainties that we set at the level of $7\%$ (CCQE $\nu_e$ and $\overline\nu_e$),
9\% (background $\nu_e$ and $\overline\nu_e$), 7\% (CCQE $\nu_\mu$ and $\overline\nu_\mu$ background and signal events), and
20\% (CC$1\pi$)
A likelihood function $L$ including Poisson statistics \cite{Cowan} is then constructed and converted into 
$\chi^2=-2\log(L)$. 

For NOvA \cite{nova_2018}, disappearance data include 126 $\nu_\mu$ events, divided into 4 subsets called 
``quantiles,'' each of them corresponding to different energy resolutions (6, 8, 10 and 12 \%)
and to 18 bins with different width. Appearance data include 57 $\nu_e$ events divided into 6 equally-spaced bins
in the interval $[1,\,4]$~GeV; we do not consider a further separation into three subclasses with different
values of the particle identification (PID) parameter \cite{nova_2018}. Also included are 9 so-called peripheral $\nu_e$ events 
grouped in a single bin. As for T2K, also NOvA backgrounds are assumed to be oscillation-independent,
oscillated spectra are tuned at best fit, and nuisance parameters are slightly adjusted. 
In particular, we assume 10\% normalization error for $\nu_e$ background and signal events, and 20\% and 5\% errors for the 
normalization and calibration of $\nu_\mu$ events, respectively. A Poissonian $\chi^2$
is then constructed.

We obtain very good agreement with all the oscillation parameter constraints shown
in \cite{t2k_2017} for T2K and in \cite{nova_2018} for NOvA,  under the same assumptions used therein about $\theta_{13}$ (usually
restricted around $\sin^2\theta_{13}\simeq 0.02$).  
We emphasize that no restrictive assumption is made when the T2K and NOvA  data   are included 
in the global analysis, all the parameters being left free. We also remark that T2K and NOvA appearance data are now
accurate enough to require analyses in terms of binned spectra rather than of total rates, the 
latter being less constraining: this represents tremendous progress in the field. 
Finally, it should be noticed that  the two collaborations
are working towards the formation of a joint analysis group producing a full T2K+NOvA combined analysis 
by 2021 \cite{T2K+NOVA}.  

\newpage
We conclude by reminding that the $3\nu$ appearance probability of
accelerator neutrinos (traveling along a baseline $x$ in constant $N_e$)  can be approximately cast in the form 
(in natural units) \cite{Petcov,Cervera:2000kp,Freund:2001pn}:

\vspace*{-8pt}
\footnotesize
%------------------------------
\begin{eqnarray}
P({\nu_\mu\to\nu_e}) & \simeq & 
\sin^2\theta_{23}\sin^22\theta_{13}\left(\frac{\Delta m^2}{A-\Delta m^2}\right)\sin^2\left(\frac{A-\Delta m^2}{4E}x\right)\nonumber\\
&+& \sin2\theta_{23}\sin2\theta_{13}\sin2\theta_{12}\left(\frac{\Delta m^2}{A}\right)
\left(\frac{\Delta m^2}{A-\Delta m^2}\right)\sin\left(\frac{A}{4E}x\right)\sin\left(\frac{A-\Delta m^2}{4E}x\right)\cos\left(\frac{\Delta m^2}{4E}x\right)
\cos\delta \nonumber\\
&-& \sin2\theta_{23}\sin2\theta_{13}\sin2\theta_{12}\left(\frac{\Delta m^2}{A}\right)
\left(\frac{\Delta m^2}{A-\Delta m^2}\right)\sin\left(\frac{A}{4E}x\right)\sin\left(\frac{A-\Delta m^2}{4E}x\right)\sin\left(\frac{\Delta m^2}{4E}x\right)\sin\delta \nonumber\\
&+& \cos^2\theta_{13}\sin^22\theta_{12}\left(\frac{\Delta m^2}{A}\right)^2
\sin^2\left(\frac{A}{4E}x\right)\ ,
\label{Pacc}
\end{eqnarray}
%------------------------------
\normalsize
where $A=2\sqrt{2}G_F N_e E$ governs matter effects, with $A\to-A$ and $\delta\to-\delta $ for $\nu\to\overline\nu$, and $\Delta m^2\to-\Delta m^2$ for
normal to inverted ordering. At typical NOvA energies ($E\sim 2$~GeV) it is $|A/\Delta m^2|\sim 0.2$, and significant matter effects 
can build up along the baseline $x=810$~km. At the lower T2K energies, the ratio $|A/\Delta m^2|$ is a factor 3--4 smaller, and
the baseline ($x=295$~km) is also smaller, so that oscillations are almost vacuum-like. See also
\cite{Akhmedov:2004ny,Takamura:2004cr,He:2016dco,Li:2016pzm,Parke:2018brr} for recent analytical studies of $3\nu$ probabilities at accelerator baselines and energies.
Note that the above form for $P_{\mu e}$, despite being useful for later discussions, is not used in our analysis, which
is based on full $3\nu$ numerical probabilities in matter
(for both appearance and disappearance channels) without any approximation.

\subsection{Short-baseline reactor neutrinos}

At the time of \cite{Fogli:2005cq}, short-baseline reactor neutrino results \cite{Apollonio:2002gd} 
were compatible with null oscillations within
statistical and systematic errors. The development of the near-far detection technique 
\cite{Martemyanov:2002td} and the construction 
of massive detectors allowed to reduce the uncertainties and to discover $\theta_{13}$, 
currently measured by three experiments: Daya Bay \cite{An:2012eh,An:2015nua,An:2016ses}, 
RENO \cite{Ahn:2012nd,RENO:2015ksa,Seo:2017ksq} and Double Chooz \cite{Abe:2011fz,Abe:2014bwa,GilBotella:2017dxg}; 
see \cite{Kim:2013vda,Vogel:2015wua,Qian:2018wid} for recent reviews. Detailed spectral
information actually allows to determine joint bounds on $(\Delta m^2,\,\theta_{13})$, as demonstrated
by RENO \cite{Seo:2016uom} and Daya Bay \cite{An:2016ses}, the latter setting bounds on $\Delta m^2$ 
competitive with those from LBL accelerator data \cite{Qian:2018wid}. 
These results represent a major success of reactor neutrino physics.

Among reactor experiments, Daya Bay \cite{An:2016ses} dominates current bounds on $(\Delta m^2,\,\theta_{13})$, the corresponding 
uncertainties being a factor $\sim 2.5$ smaller than in RENO \cite{Seo:2016uom,Seo:2017ksq} and significantly smaller than in Double Chooz 
\cite{GilBotella:2017dxg} (see also \cite{Qian:2018wid}). 
Recent analyses have shown that the reactor data combination only leads to
fractional differences (well below $1\sigma$) in comparison with bounds from 
Daya Bay data alone \cite{Esteban:2016qun,deSalas:2017kay}. 

Systematic errors are already comparable to statistical ones in both Daya Bay and RENO,
and some systematics are shared by all reactor experiments including Double Chooz. Therefore, a proper combination should
take into account a common set of nuisance parameters affecting the three experiments at the same time,
within a unified analysis framework. A joint analysis  might possibly clarify the apparent
preference of Double Chooz for higher
values of $\theta_{13}$ \cite{Qian:2018wid} as compared with Daya Bay and RENO. 
Work is in progress towards this (technically difficult but scientifically
worthwhile) joint analysis, as testified by dedicated meetings \cite{Meeting1,Meeting2} and ongoing common activities
mentioned, e.g., in \cite{Mention1,Mention2,Mention3}. For the purposes of this work, lacking
a full understanding of common systematics, we choose to limit ourselves to using 
the official $\chi^2$ map from Daya Bay alone in the global analysis.
Such a Daya Bay map is provided in terms of 
$\chi^2=\chi^2(\Delta m^2_{ee},\,\sin^2\theta_{13})$ \cite{daya_bay_chi2_map}, where the effective parameter $\Delta m^2_{ee}$ \cite{Parke:2016joa}
can be converted into $\Delta m^2$ via the relation \cite{Capozzi:2013psa,Capozzi:2015bpa}
%------------
\begin{equation}
\Delta m^2_{ee} = |\Delta m^2|\pm (c^2_{12}-s^2_{12})\delta m^2/2\ ,
\end{equation}
%--------------
where the upper (lower) sign refers to NO (IO). A proper combination of data and correlated uncertainties from all three reactor experiments 
is left as a future opportunity.

\subsection{Atmospheric neutrinos}
\medskip

Atmospheric neutrinos represent a very important and rich source of information on neutrino oscillations, 
culminating in the discovery of $\nu_\mu\to\nu_\mu$ disappearance driven by $(\Delta m^2,\,\theta_{23})$ in 1998 \cite{Fukuda:1998mi,Kajita:2000mr}.
The wide range of energies and baselines probed by atmospheric neutrinos and antineutrinos of both muon and electron
flavors, 
makes them sensitive to interesting multi-layer matter effects \cite{Petcov:1998su,Akhmedov:1998ui,Akhmedov:1998xq,Chizhov:1999he}
and to all the oscillation parameters in the $3\nu$ framework \cite{Akhmedov:2006hb,Akhmedov:2008qt,Choubey:2016gps}, 
although only
the dominant ones $(\Delta m^2,\,\theta_{23})$ have been really measured (with stringent
upper and lower bounds) within this data sample so far \cite{Kajita2016}. 

\medskip
In general, the event rate $R_\beta$ for lepton-like events of flavor $\beta =e,\,\mu$ induced by atmospheric neutrinos of the
same ($\beta$) or different ($\alpha$) flavor must be estimated through multi-dimensional integrals of the form 
\cite{Fogli:1996nn,Fogli:1998au,GonzalezGarcia:2007ib,Ge:2013ffa}
%----------------
\begin{equation}
R_\beta = \int (\Phi_\beta P_{\beta\beta} + \Phi_\alpha P_{\alpha\beta}) \otimes \sigma _\beta \otimes r_\beta \otimes \varepsilon_\beta\ , 
\end{equation}
%------------------  
where $\Phi$ represents the initial neutrino fluxes, $P$ the oscillation probability, $\sigma$ the cross section,
and $r$ and $\varepsilon$ the detection resolution for the final-state lepton, while $\otimes$ generically denotes convolution.
The rates $R_{\beta}$ are usually subdivided according to specific event topologies to specific ranges in observed energy
and angular (bins). Finite energy and direction resolutions smear out considerably the information in $P$ over binned
spectra.
All these ingredients come with their own uncertainties, which need to be estimated and propagated 
to the various spectra, often inducing sample-to-sample and bin-to-bin correlations of systematics \cite{Fogli:2003th,GonzalezGarcia:2007ib}. 
See also \cite{Capozzi:2015bxa,Capozzi:2017syc} for statistical issues in the analysis of prospective data from future large-volume atmospheric neutrino detectors.
The whole analysis is quite sophisticated 
and is becoming increasingly difficult---if not impossible---to be constructed outside the experimental collaborations
themselves.

\medskip
For instance, the latest Super-Kamiokande (SK) atmospheric data samples include as many as 
520 bins in energy-angle and 155 systematic parameters \cite{Abe:2017aap}, whose complete analysis 
vastly exceeds the capabilities of any external user. In particular, recent techniques for
the statistical separation of $\nu_e$ and $\overline\nu_e$ event in dedicated samples \cite{Abe:2017aap}, which
provide enhanced  sensitivity to matter effects, neutrino CP violation and mass ordering, has been achieved
via neural-network simulations of the detection process \cite{Tristan2014}. Another issue is represented by
incomplete (or missing) public information. As an example, the IceCube DeepCore (IC-DC)
atmospheric data release in \cite{Aartsen:2014yll} was accompanied by a public analysis toolkit \cite{deepcore_chi2_map_2015}, but such tools have not (yet) been provided for 
the latest data release in \cite{Aartsen:2017nmd},
preventing a direct use by external users. This drawback
has been recently compensated by the availability of IC-DC $\chi^2$ maps 
\cite{deepcore_chi2_map_2018} derived from the official oscillation analysis in \cite{Aartsen:2017nmd}.

\medskip
There may be different approaches to these issues, especially concerning the vast and complex SK data set. 
One may limit the analysis to those subsets of SK atmospheric
data which can be reliably reproduced outside the collaborations, as it was attempted in most global analyses so far,
including, e.g., \cite{Fogli:2005cq,Fogli:2012ua,Capozzi:2016rtj,GonzalezGarcia:2012sz,GonzalezGarcia:2007ib,Schwetz:2008er}. 
Alternatively, one may use official $\chi^2$ maps from SK if available, as advocated in \cite{Tortola:2012te,deSalas:2017kay} 
that included the results in \cite{Wendell:2010md,SK_old_chi2_map}. Note, however, that 
such maps were obtained in the one-dominant mass-scale approximation ($\delta m^2=0$) \cite{Wendell:2010md,SK_old_chi2_map}
and thus, by construction, they are insensitive to several subleading $3\nu$ effects, including CP violation. 
Eventually---and more radically---one
 may just ``give up'' on the analysis of SK atmospheric data and exclude them altogether, as recently advocated 
in \cite{Esteban:2016qun,nufit}. 

\newpage
In this work, we prefer to
abandon our own analysis  of SK atmospheric data, improved over the last twenty years 
\cite{Fogli:1998au,Capozzi:2017ipn}.
We feel that our attempts to analyze 
these data are no longer competitive with the official SK ones, as unavoidable approximations 
and data selections might bias or 
hinder the emergence of small, subleading effects. However, it makes sense to keep such data in the global analysis, since
the full SK atmospheric sample clearly shows an increasing sensitivity
to several $3\nu$ ``unknowns'' \cite{Abe:2017aap}, especially in combination with reactor and accelerator data.
We thus adopt
the official SK $\chi^2$ maps which have been recently available \cite{SK_new_chi2_map}, as derived in \cite{Abe:2017aap}
through a full $3\nu$ analysis of atmospheric data only (without external constraints from reactor 
or accelerator experiments). These SK $\chi^2$ maps are provided for both NO and IO in terms 
of four relevant parameters $(\Delta m^2,\,\theta_{23},\,\theta_{13},\,\delta)$,
with $(\delta m^2,\,\theta_{12})$ fixed at best fit.  We also adopt the maps provided by 
IC-DC for their latest data sets \cite{Aartsen:2017nmd,deepcore_chi2_map_2018}
in terms of the two dominant parameters $(\Delta m^2,\,\theta_{23})$ for both NO and IO,
with $(\delta m^2,\,\theta_{12})$ fixed at best fit, the dependence on $\theta_{13}$ and $\delta$ being
negligible at the current level of accuracy in IC-DC data \cite{Private}.
%---------------------- 
%\footnote{
%We do not include  data from other atymospheric neutrino experiments,
%as they would induce negligible changes in the parameter estimates, with respect to the dominant constraints
%from SK and IC-DC.}

We conclude this section by discussing the context and implications of this choice.
Atmospheric neutrinos provide free beams 
with wide dynamical range in energy and pathlength, which will always provide a vast amount
of interesting data (signal or background) in underground detectors. In particular,
they still contain very rich oscillation physics to be explored, 
especially in terms of subleading $3\nu$ effects \cite{Akhmedov:2008qt,SubWork,Subdominant} that, however, can emerge only
through increasingly sophisticated analyses. Probably only the experimental
collaboration are (and will be) able to study their own atmospheric data at such a refined level.
This transition should be accompanied by 
a continuous dialogue with the scientific  community, in order to make 
progress on several ``unknowns'' or poorly known quantities that may hide
the relevant atmospheric neutrino physics, including e.g., the systematics of cosmic ray fluxes, 
atmosphere parameters, cascade evolution models,  three-dimensional effects, 
event spectral shapes, resolution tails, effective volume estimates, detection cross sections, etc.
The renewed interest in these issues is  testified by recent dedicated atmospheric $\nu$ workshops \cite{ANW16,PANE18}, 
in addition to traditional series with broader scope \cite{NEUTEL,VLVNT}. Such topics
will become even more crucial  in the future, 
to make the best possible use of high-statistics data 
coming from new-generation projects \cite{Yanez:2015uta} such as PINGU \cite{Aartsen:2014oha,TheIceCube-Gen2:2016cap},
KM3NeT-ORCA \cite{Adrian-Martinez:2016fdl}, Hyper-Kamiokande \cite{Abe:2011ts} and INO \cite{Athar:2006yb}.

In summary, we continue to perform an independent analysis of solar and KamLAND data,
as well as of long-baseline accelerator neutrino data (with updated results from T2K \cite{t2k_2017} and NOvA \cite{nova_2018}), 
whose combination provides
bounds on the whole set of $3\nu$ oscillation parameters $(\delta m^2,\,\Delta m^2,\,\theta_{12},\,\theta_{13},\,\theta_{23},\,\delta)$.
For the first time, we use processed results rather than original data for the other data sets.
In particular, we include short-baseline reactor data constraints via the official Daya Bay map 
$\chi^2(\Delta m^2,\,\theta_{13})$ \cite{daya_bay_chi2_map}, and finally atmospheric neutrino data via
the official SK map $\chi^2(\Delta m^2,\,\theta_{23},\,\theta_{13},\delta)$
\cite{SK_new_chi2_map} and the
IC-DC map $\chi^2(\Delta m^2,\,\theta_{23})$ \cite{deepcore_chi2_map_2018}. We have argued that
using (some) processed results from experimental collaborations is becoming unavoidable
in global analyses, although this transition should be accompanied by critical discussions
and further advances in several related subfields. We have limited ourselves to the $3\nu$ paradigm, but the
same arguments apply to the analysis of possible subleading effects coming from
extended frameworks with new neutrino states or interactions.

%%%%%%%%%%%%%%%%%%%%%%%%%%%%%%%%%%%%%%%%%%%%%%%%%%%%%%%%%%%%%%%%%%%%%%%%%%%%%%%%%%%%%%%%%%%%%%%%%%%%%%%%%
\section{Results on single oscillation parameters}

In this section we present the bounds on known and unknown $3\nu$ oscillation parameters, coming
from the data sets discussed in the previous section. Bounds are shown in terms of single parameters,
all the others being marginalized away. The discussion of some 
detailed features, that involve the interplay
between different parameters (covariances) is postponed to Section~4. The main new result is the 
emerging indication in favor of NO at $\sim 3\sigma$ in the global analysis, 
with coherent contributions from all data sets.
We also briefly compare our results with those
obtained in other recent analyses \cite{Esteban:2016qun,nufit,deSalas:2017kay} under homogeneous assumptions as far as possible. 

%\newpage
\subsection{Synopsis with increasingly large data sets}

{\bf \em Analysis of long-baseline accelerator, solar and KamLAND data.} 
Figure~1 shows the bounds on single oscillation parameters, in terms of standard deviations $N\sigma$ from 
the best fit, for both NO (solid blue lines) and IO (dashed red lines), with separate $\chi^2$ minimization for
the two mass orderings. Symmetric and linear curves would correspond to gaussian errors, a situation approximately
realized for the parameters $\delta m^2$, $\theta_{12}$ and $\Delta m^2$. Strong upper and lower bounds 
are placed on the $\Delta m^2$, $\theta_{23}$ and $\theta_{13}$ parameters.
Thus, the combination of long-baseline accelerator, solar and KamLAND data  provides, by itself, a measurement of
the known oscillation parameters. In addition, interesting hints emerge on the unknown ones.

Concerning  the phase $\delta$, the CP-conserving values 
$\delta=\{0,\,\pi\}$ are allowed at $\sim 2\sigma$ or less in both NO and IO. However, there is a clear preference for
values around $\delta\sim 3\pi/2 $, i.e.\ for nearly maximal CP violation with $\sin\delta\sim -1 $, while 
values near the opposite case with $\sin\delta\sim +1 $ are disfavored at more than $3\sigma$. Concerning
the octant of $\theta_{23}$, there is a slight preference for $\theta_{23}<\pi/4$ in NO and
$\theta_{23}>\pi/4$ in IO, but both octants are allowed at $1\sigma$.

\begin{figure}[t]
\begin{center}
\begin{minipage}[t]{16.5 cm}
\center
\epsfig{file=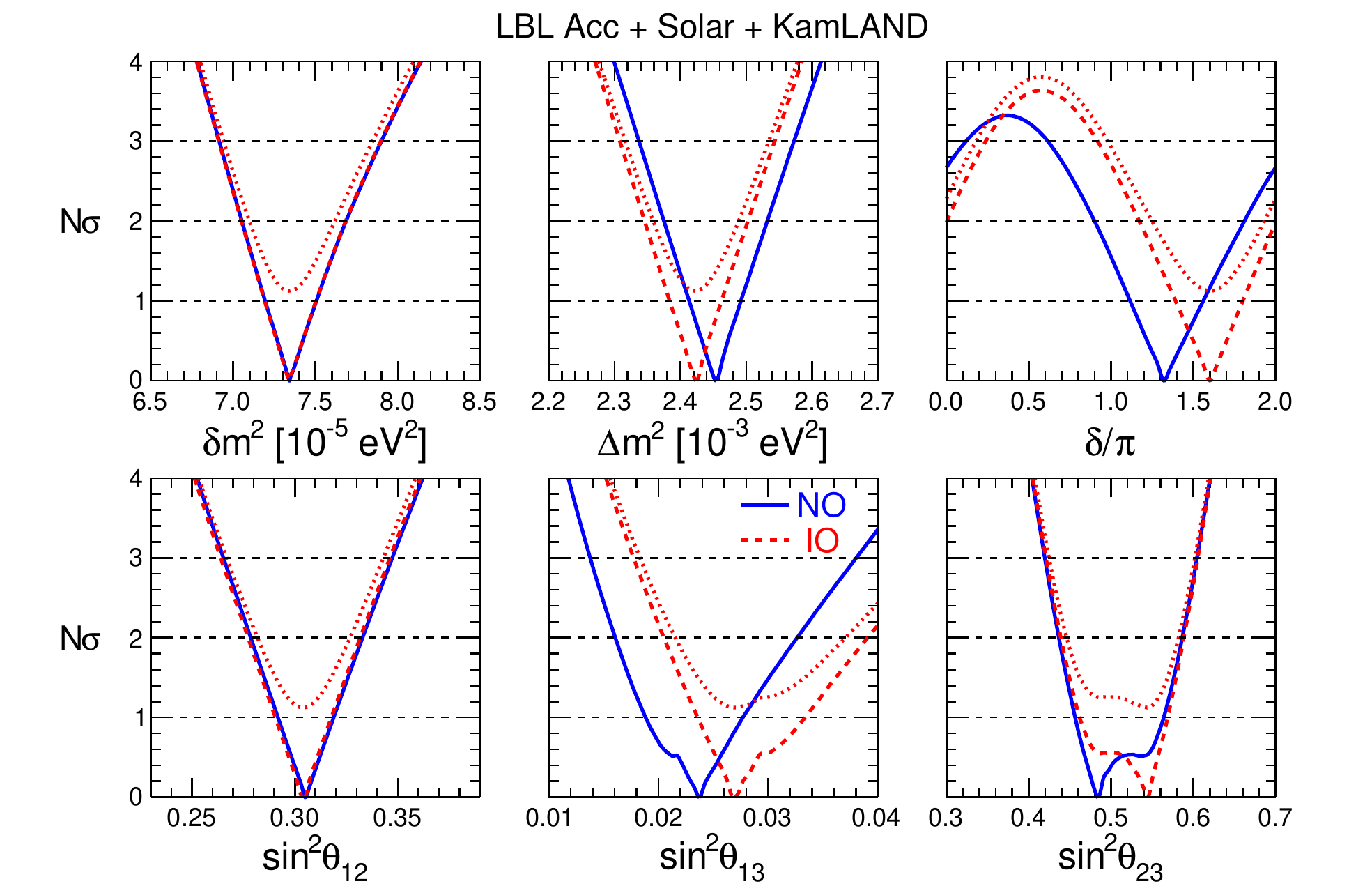,scale=0.71}
\end{minipage}
\begin{minipage}[t]{16.5 cm}
%\vspace*{-3mm}
\caption{
Analysis of long-baseline accelerator, solar and KamLAND data. Bounds on the mass-mixing parameters
are given in terms of standard deviations $N\sigma =\sqrt{\chi^2-\chi^2_{\min}}$ for both normal ordering (NO, solid blue lines)
and inverted ordering (IO, dashed red lines), taken  separately. 
For IO, bounds are also shown with respect to the absolute $\chi^2_{\min}$ for NO (dotted red curves). 
The IO is slightly disfavored, at the level of $N\sigma \simeq  1.1$.   
\label{fig_01}}
\end{minipage}
%\vspace*{-3mm}
\end{center}
\end{figure}

Concerning the mass ordering, Fig.~1 shows that the
bounds on both $\delta m^2$ and $\theta_{12}$ (dominated by solar+KL data) are almost completely insensitive to it. 
On the contrary, some differences are found between NO and IO
for the best-fit values and allowed ranges of the parameters
$(\Delta m^2,\, \theta_{23},\, \theta_{13},\, \delta)$, that are constrained by 
 long-baseline accelerator data. In particular, there is a preference for higher $\theta_{13}$ in IO.
We also find an overall difference between the two $\chi^2$ minima in NO and IO, that amounts to
%-----------
\begin{equation} 
\chi^2_{\min}(\mathrm{IO})-\chi^2_{\min}(\mathrm{NO})=1.3\  \
\mathrm{(LBL\ acc.+solar+KL\ data)}\ , 
\label{chi1}
\end{equation}
%------------
corresponding to a slight preference 
for NO at the level of $N\sigma\simeq 1.1$.  

We thus show in Fig.~1 the parameter bounds for IO
in terms of $N\sigma$, also by taking into account the absolute minimum in NO  (dotted red lines). 
For any parameter in Fig.~1, marginalization over the (unknown) hierarchy information would 
correspond to taking the union of the allowed ranges at some $N\sigma$ 
for NO (blue solid curves) and for the displaced IO ones (dotted red curves). 

The $\Delta\chi^2$ difference in Eq.~(\ref{chi1}) is almost entirely driven by the recent T2K and NOvA data.
Older K2K and MINOS data are less relevant, and actually their removal would lead to a slightly higher 
preference for NO (not shown). Further T2K and NOvA results, possibly combined by the Collaborations
themselves \cite{T2K+NOVA}, will be crucial to test the current trend favoring NO over IO in this data sample.

\begin{figure}[t]
\begin{center}
\begin{minipage}[t]{16.5 cm}
\center
\epsfig{file=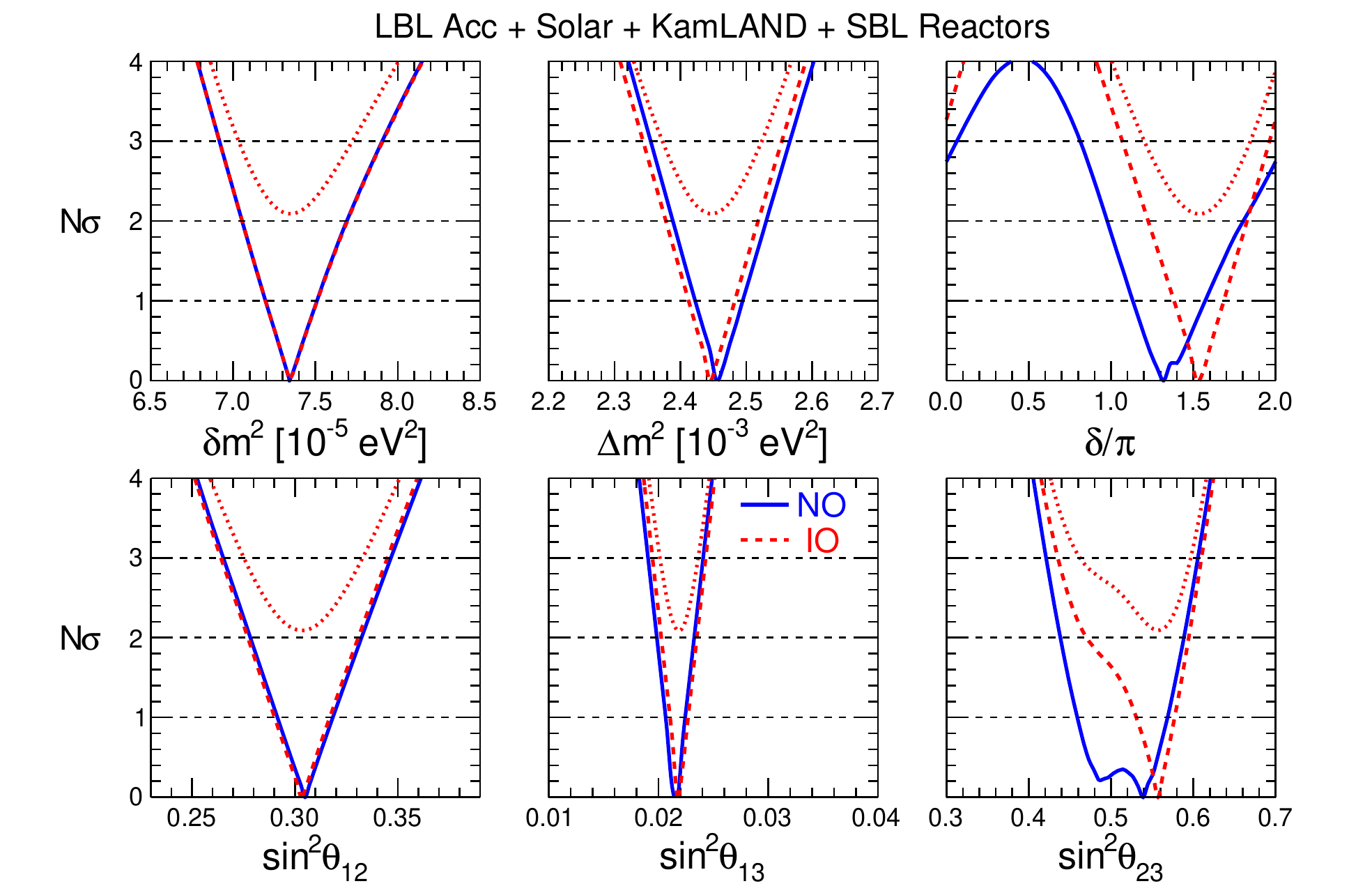,scale=0.71}
\end{minipage}
\begin{minipage}[t]{16.5 cm}
\caption{Analysis of long-baseline accelerator, solar and KamLAND data, and short-baseline
reactor data. Line styles and colors are as in Fig.~1.    
\label{fig_02}}
\end{minipage}
\end{center}
\end{figure}

\medskip

{\bf \em Adding short-baseline reactor data.} 
Figure~2 is analogous to Fig.~1, but includes short-baseline reactor constraints as 
described in Section~2. With respect to Fig.~1, 
the allowed range for $\theta_{13}$ is strongly reduced, with nearly linear and symmetric bounds 
for both NO and IO.  Also the allowed range for $\Delta m^2$ is noticeably reduced, showing that
reactor neutrinos are already competitive with long-baseline accelerators in determining the
largest oscillation frequency driven by $\Delta m^2$. Both parameters $(\Delta m^2,\,\theta_{13})$
depend much less on the mass ordering than in Fig.~1.   

Concerning the unknown parameters, the octant ambiguity of $\theta_{23}$ remains unresolved, but there is a mild overall preference for 
$\theta_{23}>\pi/4$, more pronounced for IO.
The indications in favor of nearly maximal CP violation are instead strengthened, and the CP-conserving
values of $\delta$ are now disfavored at the level of $>1.8\sigma$ in NO and $>3\sigma$ in IO.
Significant ranges for $\delta$ are excluded at $>3\sigma$ in both NO and IO.  
 The preference for NO is also corroborated, and amounts to  
%-----------
\begin{equation} 
\chi^2_{\min}(\mathrm{IO})-\chi^2_{\min}(\mathrm{NO})=4.4\  \
\mathrm{(LBL\ acc.+solar+KL+SBL\ reac.\ data)}\ , 
\label{chi2}
\end{equation}
%------------
corresponding to an interesting confidence level $N\sigma\simeq 2.1$. 
As discussed in more detail in Sec.~4, the above result stems mainly from a slight $\theta_{13}$ tension in IO 
between reactor and accelerator data, the latter preferring higher values of $\theta_{13}$ than the former.

\begin{figure}[t]
\begin{center}
\begin{minipage}[t]{16.5 cm}
\center
\epsfig{file=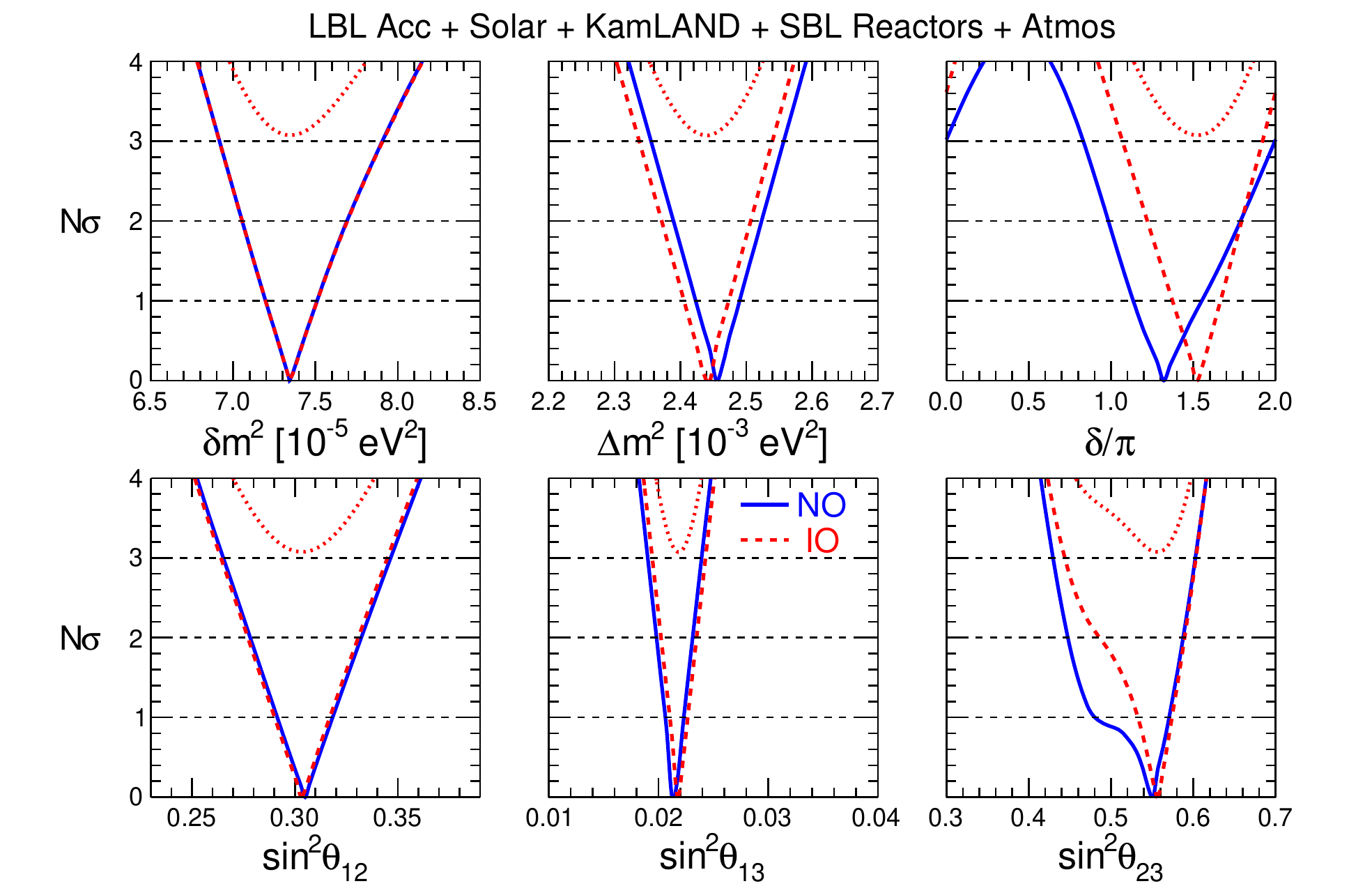,scale=0.71}
\end{minipage}
\begin{minipage}[t]{16.5 cm}
\caption{Global analysis of oscillation data from long-baseline accelerator, solar and KamLAND, short-baseline
reactor, and atmospheric neutrino experiments. Line styles and colors are as in Fig.~1.    
\label{fig_03}}
\end{minipage}
\vspace*{-4mm}
\end{center}
\end{figure}

{\bf \em Adding atmospheric neutrinos: Global analysis of all oscillation data.}
Figure~3 is analogous to Fig.~2, but includes atmospheric neutrino constraints 
as described in Section~2. With respect to Fig.~2, the main differences concern the unknown 
oscillation parameters. There is a more pronounced preference for 
$\theta_{23}>\pi/4$, although both octants are allowed at $<2\sigma$. The preference 
for CP violation with $\sin\delta<0$ is confirmed, while CP conservation is disfavored at
$>1.9\sigma$ for NO and $>3.5\sigma$ for IO.  Remarkably, 
the sensitivity of atmospheric data to the mass ordering is also consistent with the hints from 
previous data sets and leads to 
%-----------
\begin{equation} 
\chi^2_{\min}(\mathrm{IO})-\chi^2_{\min}(\mathrm{NO})=9.5\ \ \mathrm{(all \ oscillation\ data)}\ , 
\label{chi3}
\end{equation}
%------------
corresponding to a statistically significant confidence level $N\sigma\simeq 3.1$. 
The increase from Eq.~(\ref{chi2}) to Eq.~(\ref{chi3}) is mainly due to SK atmospheric data
\cite{Abe:2017aap}, but there is also a synergic contribution (by about one unit of $\Delta\chi^2$) 
from IC-DC data, that will be discussed in Sec.~4.

\subsection{Summary and discussion of results}

The preference for NO at the level of $\Delta\chi^2\sim 9 $ in Eq.~(\ref{chi3}) represents an interesting result of our work.
This indication emerges consistently 
for increasingly rich data sets, as shown by the progression in 
Eqs.~(\ref{chi1})--(\ref{chi3}), and thus deserves attention. 
Taken at face value, a $3\sigma$ rejection of IO would imply that the only
relevant scenario is NO, together with its parameter ranges (see Fig.~3).
 
However, caution should be exercised at this stage, since
the value $\Delta\chi^2 \sim 9$ derives from two main contributions of comparable size $\Delta\chi^2\simeq 4$--5 (corresponding to $\sim 2\sigma$)
but with rather different origin. One contribution [Eq.~(\ref{chi2})] comes basically from 
long-baseline accelerator data and their interplay with short-baseline reactor data, where mass-ordering effects can be understood
with relatively simple arguments in terms of $\theta_{13}$ (see next Section). The other incremental contribution [from Eq.(\ref{chi2}) to (\ref{chi3})] 
comes basically from atmospheric data, where mass-ordering effects are not
apparent ``at a glance'', but are indirect and largely smeared over various energy-angle spectra 
\cite{Abe:2017aap}.
A wise attitude is to wait for further data from all the running experiments which, 
in the the next few years, can reveal if these two hints at $\sim 2\sigma$ level  
will fluctuate down, or will consistently grow and  confirm 
the preference for NO at a cumulative level $>3\sigma$. On a longer time frame, discovery-level tests of the mass spectrum ordering 
will be provided by next-generation projects \cite{Patterson:2015xja}, not only with large-volume
atmospheric neutrinos \cite{TheIceCube-Gen2:2016cap,Adrian-Martinez:2016fdl,Abe:2011ts,Athar:2006yb} 
but also with medium-baseline reactors such as JUNO \cite{An:2015jdp} 
and new long-baseline accelerator facilities such as T2HK \cite{Abe:2015zbg}, DUNE \cite{Strait:2016mof} and ESSnuSB \cite{Wildner:2015yaa}.
Finally, for discrete hypotheses like NO versus IO, the statistical interpretation of $\Delta\chi^2$  in terms of $N\sigma$ 
remains effectively applicable, but must 
be taken with a grain of salt \cite{Blennow:2013oma}.

In the following, we shall thus 
conservatively report the allowed ranges for NO and IO as if they were two separate and equally acceptable cases,
without including the large $\chi^2$ difference of IO with respect to the absolute minimum in NO [Eq.~(\ref{chi3})].
Marginalization over ``any ordering'' (as performed, e.g., in \cite{Esteban:2016qun,nufit,Capozzi:2017ipn}) is not considered herein.

%-----------------------------------------------------------------------------------------------
\begin{table}
\begin{center}
\begin{minipage}[t]{16.5 cm}
\caption{\small Best fit values and allowed ranges at $N\sigma=1,\,2,\,3$ for the $3\nu$ oscillation parameters, in either
NO or IO. The latter column shows the formal ``$1\sigma$ accuracy'' for each parameter, defined as $1/6$ of the $3\sigma$ 
range divided by the best-fit value (in percent). }
\label{tab:1}
%\vspace*{3mm}
\end{minipage}
\begin{tabular}{lcccccc}
\hline
\hline
Parameter & Ordering & Best fit & $1\sigma$ range & $2\sigma$ range & $3\sigma$ range & ``$1\sigma$'' (\%)\\
\hline
$\delta m^2/10^{-5}$~eV$^2$ & NO  & 7.34 & 7.20 -- 7.51 & 7.05 -- 7.69 & 6.92 -- 7.91 & 2.2\\
 							& IO  & 7.34 & 7.20 -- 7.51 & 7.05 -- 7.69 & 6.92 -- 7.91 & 2.2 \\
\hline
$\sin^2\theta_{12}$			& NO  & 3.04 & 2.91 -- 3.18 & 2.78 -- 3.32 & 2.65 -- 3.46 & 4.4 \\
							& IO  & 3.03 & 2.90 -- 3.17 & 2.77 -- 3.31 & 2.64 -- 3.45 & 4.4\\
\hline							
$\sin^2\theta_{13}/10^{-2}$ & NO  & 2.14 & 2.07 -- 2.23 & 1.98 -- 2.31 & 1.90 -- 2.39 & 3.8 \\
							& IO  & 2.18 & 2.11 -- 2.26 & 2.02 -- 2.35 & 1.95 -- 2.43 & 3.7 \\
\hline
$|\Delta m^2|/10^{-3}$~eV$^2$&NO  & 2.455 & 2.423 -- 2.490 & 2.390 -- 2.523 & 2.355 -- 2.557 & 1.4\\
							& IO  & 2.441 & 2.406 -- 2.474 & 2.372 -- 2.507 & 2.338 -- 2.540 & 1.4 \\
\hline
$\sin^2\theta_{23}/10^{-1}$ & NO  & 5.51 & 4.81 -- 5.70 & 4.48 -- 5.88 & 4.30 -- 6.02 & 5.2\\
							& IO  & 5.57 & 5.33 -- 5.74 & 4.86 -- 5.89 & 4.44 -- 6.03 & 4.8 \\
\hline
$\delta/\pi$ 				& NO  & 1.32 & 1.14 -- 1.55 & 0.98 -- 1.79 & 0.83 -- 1.99 & 14.6\\
							& IO  & 1.52 & 1.37 -- 1.66 & 1.22 -- 1.79 & 1.07 -- 1.92 & 9.3\\
\hline
\hline
\end{tabular}
\end{center}
\vspace*{-5mm}
\end{table}

Table~1 reports in numerical form {what is} shown in Fig.~3 for NO and IO separately, in terms of 
best-fit values and allowed ranges at $N\sigma=1,\,2,\,3$ level. The last column reports the fractional
``$1 \sigma$'' accuracy, defined as $1/6$ of the $3\sigma$ range, divided by the best-fit value. 
From top to bottom, the rows of Table~1 provide information on both known and unknown $3\nu$ parameters.  
The two known parameters $\delta m^2$ and $\sin^2\theta_{12}$, 
dominated by solar and KamLAND data, are basically the same in NO and IO,  
up to tiny variations discussed later in Sec.~4. They are determined 
with an accuracy of 2.2 and 4.4\%, respectively. The value of $\sin^2\theta_{13}$, dominated by
reactor data, is also determined with a very good accuracy of $\sim 3.8\%$. 
The best known
parameter is $|\Delta m^2|$, which is currently determined with a remarkably small uncertainty
of 1.4\% in each of the two mass orderings, as a result of consistent and comparable constraints from 
long-baseline accelerator, reactor and atmospheric neutrino data.

The {least} accurate among the known oscillation parameters in Table~1 is
$\sin^2\theta_{23}$, with an uncertainty of $\sim5\%$. As compared with previous analyses \cite{Capozzi:2016rtj,Capozzi:2017ipn},
the $\theta_{23}$ uncertainty is smaller, as a result of a better convergence of
recent NOvA data \cite{nova_2018} towards quasi-maximal mixing, 
in agreement with T2K and atmospheric data. Maximal mixing is allowed  at $<2\sigma$  in both NO and IO,
and thus the octant degeneracy remains unsolved. 
Concerning $\delta$, Fig.~3 shows that there is a single 
$3\sigma$ range around its best fit, with relatively linear and symmetric
errors, in both NO and IO. Nonlinear errors, also due to the cyclic nature of 
$\delta$, emerge only at a level $> 3\sigma$ in NO and $>4\sigma$ in IO. 
Within  $<3\sigma$, 
one can tentatively say that $\delta$ is ``determined'' with an accuracy of $\sim 15\%$ ($\sim 9\%$) in NO (IO),
as reported in Table~1. 

Summarizing, the ``known'' oscillation parameters 
$\delta m^2$, $\Delta m^2$, $\sin^2\theta_{12}$, $\sin^2\theta_{13}$ and $\sin^2\theta_{23}$ are 
currently measured at the few~\% level. Concerning the ``unknown'' oscillation parameters,
interesting indications emerge in favor of NO at a global $3\sigma$ level. At the same level one can also
determine upper and lower limits for the phase $\delta$, with preference for nearly maximal CP violation. 
CP conservation is generally disfavored, 
but remains allowed at $\sim 2\sigma$ in NO. The octant of  {$\theta_{23}$} is unresolved at the 
$\sim 2\sigma$ level in both NO and IO.     
If these trends are confirmed, the mass spectrum ordering and the CP phase $\delta$ might be the first
``unknowns''  to become ``known'' (at $>3\sigma$) with further data; assessing
the octant and excluding CP conservation might instead require more effort.
 
We conclude this section by comparing our results with other recent global analyses \cite{Esteban:2016qun,nufit,deSalas:2017kay}. 
If we exclude SK atmospheric data as advocated in \cite{Esteban:2016qun,nufit}, our results agree well with theirs on both known
and unknown parameters. In particular, we obtain very similar $\chi^2$ curves for $\delta$ and for $\theta_{23}$, 
including a comparable offset $\Delta\chi^2$ between IO and NO (not shown). Given this agreement, we surmise that the authors
of \cite{Esteban:2016qun,nufit} would also obtain results very similar to ours (Fig.~3 and Table~1) by adding the 
$\chi^2$ map from the latest SK atmospheric neutrino data \cite{SK_new_chi2_map}  in their recent fit \cite{nufit}. 
Concerning  the analysis in \cite{deSalas:2017kay}, we observe qualitative agreement with
their hints on the mass ordering and the CP-violating phase. However, at the level of details,
a comparison with \cite{deSalas:2017kay} is not obvious:
 their data set  is based on earlier T2K and NOvA data, and it also includes 
the $\chi^2$ map of older SK atmospheric data \cite{SK_old_chi2_map}, that was derived in the approximation 
$\delta m^2= 0$. By construction, this approximation switches off CP-violation effects
(and may bias other subleading effects) in atmospheric neutrinos, preventing a proper and detailed comparison of global fit results.

%%%%%%%%%%%%%%%%%%%%%%%%%%%%%%%%%%%%%%%%%%%%%%%%%%%%%%%%%%%%%%%%%%%%%%%%%%%%%%%%%%%%%%%%%%%%%%%%%%%%%%%%%
\section{Results on oscillation parameter pairs}

In this Section we show the allowed regions in various planes charted by pairs 
of oscillation parameters. We discuss covariances related to $\delta m^2$-driven oscillations,
to $\Delta m^2$-driven oscillations, and to  the CP-violating phase $\delta$. We always take NO and IO
as two isolated cases, without marginalizing over the mass ordering.

\begin{figure}[hb]
\begin{center}
\begin{minipage}[t]{16.5 cm}
\center
\epsfig{file=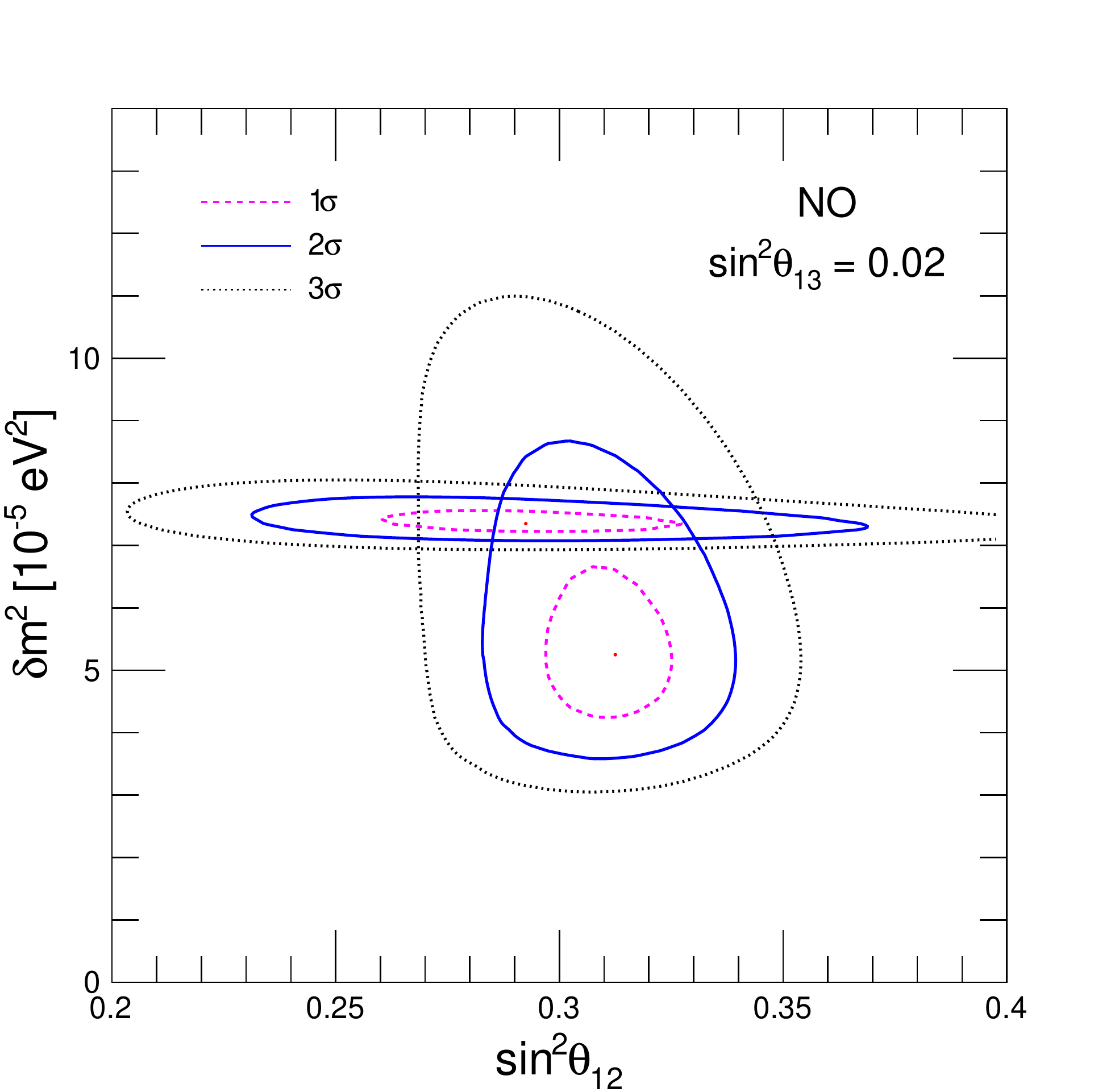,scale=0.36}
\end{minipage}
\begin{minipage}[t]{16.5 cm}
\caption{Separate analysis of solar and KamLAND neutrino experiments in the plane $(\delta m^2,\,\sin^2\theta_{12})$, 
assuming NO and a fixed value $\sin^2\theta_{13}=0.02$. The contours correspond to $N\sigma = 1,\,2,\,3$.
For IO, the solar neutrino contours would be very similar, but shifted 
by $\delta(\sin^2\theta_{12})\simeq -0.02$ (not shown). 
See the text for details.   
\label{fig_04}}
\end{minipage}
\end{center}
\end{figure}

\subsection{Covariances of $(\delta m^2,\,\theta_{12},\theta_{13})$}

The $(\delta m^2,\,\theta_{12},\theta_{13})$ parameters govern the oscillations of solar and KamLAND neutrinos,
which are of great importance not only by themselves but also for providing the $(\delta m^2,\,\theta_{12})$
input to the full $3\nu$ analysis of all the other experiments. 

Figure~4 shows the $N\sigma$ regions allowed separately by solar and KamLAND data in the plane $(\delta m^2,\,\sin^2\theta_{12})$, 
assuming NO and fixing $\theta_{13}$ at a representative value ($\sin^2\theta_{13}=0.02$). The separate results show some slight 
tension between the preferred mass-mixing values (between  $1\sigma$ and $2\sigma$ from a glance at Fig.~4), that has
received interest (see, e.g., \cite{Maltoni:2015kca,Gonzalez-Garcia:2013usa,Liao:2017awz,Ghosh:2017lim}) 
as a possible indication of nonstandard neutrino interactions in the solar matter 
(see, e.g., \cite{Palazzo:2011vg,Bolanos:2008km,Capozzi:2017auw,Coloma:2017egw,Coloma:2017ncl}). 
We remind that nonstandard four-fermion interactions with effective couplings 
$\varepsilon_{\alpha\beta}G_F$ (typically with $|\varepsilon_{\alpha\beta}|\ll 1$), may affect the precise 
determination of the oscillation parameters: in particular, flavor-diagonal couplings ($\alpha=\beta$) tend to affect
the neutrino energy differences and thus the squared mass gaps, while off-diagonal couplings  ($\alpha\neq \beta$)
may alter the mixing matrix. In the vast related literature,
see, e.g., \cite{Biggio:2009nt,Escrihuela:2009up,Ohlsson:2012kf,Miranda:2015dra,Farzan:2017xzy} 
for phenomenological reviews and \cite{Gavela:2008ra,Antusch:2008tz,Babu:2017olk} for theoretical model constructions.

We quantify the tension between the best fits in Fig.~4  in terms of the difference 
between the joint and separate $\chi^2$ minima, $\Delta \chi^2=\chi^2_\mathrm{solar +KL}-(\chi^2_\mathrm{solar}+
\chi^2_\mathrm{KL})\simeq 2$, finding $N\sigma\simeq 1.4$; this value would slightly increase to
$N\sigma =1.6$ for unconstrained $\theta_{13}$ (using only solar and KL data). We conclude that 
the overall hint in favor of nonstandard interactions in the Sun does not exceed $2\sigma$ at present. 
Should this hint be corroborated by future data, nonstandard interactions should be generally considered also in the Earth matter,
and they could affect future measurements of the known ($\Delta m^2,\,\theta_{23},\,\theta_{13}$) parameters, or perturb indications
about the unknown mass ordering.  
This possibility provides a relevant example of the interplay between the $3\nu$ oscillation parameters
and generalized ``unknowns'' coming from scenarios beyond $3\nu$, which provide an active area of research for prospective
oscillation searches; see, e.g., 
\cite{Farzan:2017xzy,Blennow:2016etl,Deepthi:2016erc,Agarwalla:2016fkh,Liao:2016orc,Fukasawa:2016nwn,Ge:2016dlx,Masud:2016gcl} 
for recent studies.

Let us now discuss the details of the $3\nu$ mass-ordering effects in Fig.~4. This figure
has been obtained for NO; in IO  the KamLAND contours
would not change, since $\Delta m^2$-driven
oscillations are effectively averaged out, and the sign of $\Delta m^2$ is not probed [see Eq.~(\ref{P3nu2nu})].  
However, for solar neutrinos,  Eq.~(\ref{P3nu2nu}) involves the effective 
$\theta_{13}$ mixing angle in matter, which embeds a slight residual dependence on  $\pm\Delta m^2$
\cite{Fogli:2001wi} and thus to mass ordering \cite{Fogli:2005cq}. We find that the solar neutrino contours in Fig.~4 would be slightly
different in IO, being basically shifted leftwards by a tiny amount, $\delta(\sin^2\theta_{12})\simeq -0.02$
(not shown). This small shift compensates, in the solar neutrino data fit, 
the slightly higher survival probability for IO as compared to NO 
(see Fig.~13 in \cite{Fogli:2001wi}).  In combination with (mass-ordering insensitive) KamLAND data, the overall shift
of the best-fit mixing angle amounts to $\delta(\sin^2\theta_{12})\simeq -0.01$ in IO, as also reported in
Table~1 for the sake of precision.
The absolute $\chi^2$ difference between the solar+KL fit amounts to a mere $\Delta\chi^2=0.08$ in favor
of IO with respect to NO; although statistically insignificant, this difference 
has been taken into account in our global analysis. 

We conclude our comments to Fig.~4 by noting that analogous results shown in \cite{Abe:2016nxk} (see Figs. 33 and 34 therein)
display solar neutrino contours with small ``wiggles'' that are not found in other
recent analyses \cite{Capozzi:2016rtj,Esteban:2016qun,nufit,deSalas:2017kay}. We surmise that
such wiggles are not physical effects (which should be smoothly varying in terms of $\delta m^2$), but 
may be numerical artifacts due to insufficient grid sampling in an integration 
variable (either energy or zenith angle).  In this specific case, an ``official'' $\chi^2$ map 
from the collaboration would not bring a clear advantage over analyses performed by external researchers, in terms of 
overall accuracy. 

\newpage

\begin{figure}[t]
\begin{center}
\begin{minipage}[t]{16.5 cm}
\center
\epsfig{file=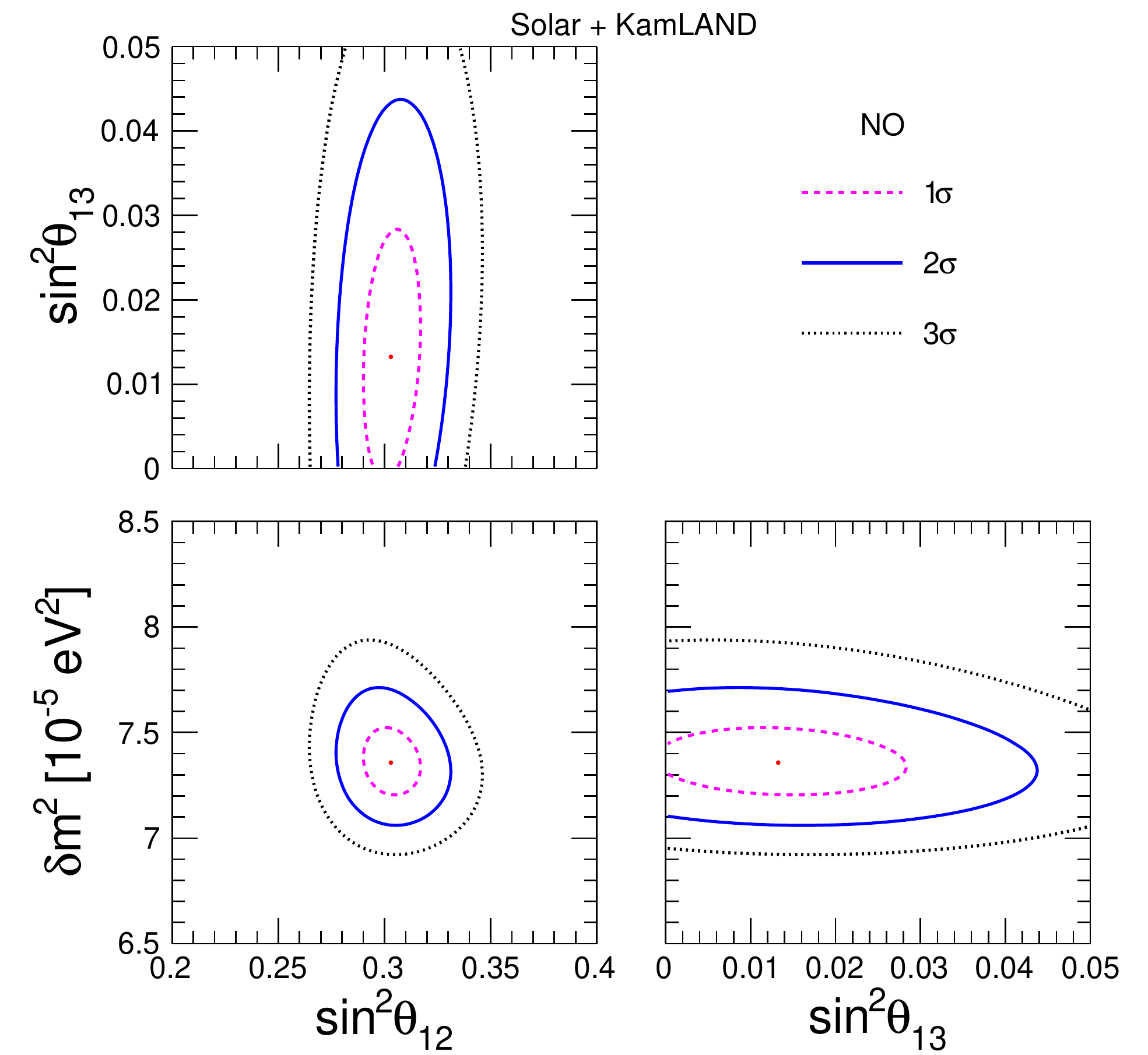,scale=0.35}
\end{minipage}
\begin{minipage}[t]{16.5 cm}
\caption{Joint analysis of solar and KamLAND neutrino data in each of the planes
charted by one pair of parameters among $(\delta m^2,\,\sin^2\theta_{12},\,\theta_{13})$. 
The contours correspond to $N\sigma = 1,\,2,\,3$. Results refer to NO, and would be very similar
for IO (not shown). See the text for details.   
\label{fig_05}}
\end{minipage}
\end{center}
\end{figure}

We now consider the covariances of any pair of parameters among $(\delta m^2,\,\theta_{12},\,\theta_{13})$,
as obtained by considering increasingly rich data sets. Figure~5 shows the covariances as obtained by solar+KL data only,
which provide the well-known weak hint $(\sim 1\sigma)$ for nonzero $\theta_{13}$ \cite{Fogli:2008jx}, with negligible correlations
of $\theta_{13}$ with the other two parameters. 

Figure~6 is analogous to Fig.~5 but includes long-baseline
accelerator data, that provide a dramatic reduction of the allowed range for $\theta_{13}$
due to flavor-appearance data. The small kink in the contours involving the $\theta_{13}$ parameter has
a physical origin, being related to the octant degeneracy of $\theta_{23}$: in a sense, the contours  
in the planes involving $\theta_{13}$ are superpositions of two regions, slightly displaced in $\theta_{13}$,
corresponding to nearly-degenerate fits in the two $\theta_{23}$ octants. See also Fig.~1, as well as the
discussion of the correlations between $\theta_{23}$ and $\theta_{13}$ in the next subsection. 
 
\vfill

\begin{figure}[hb]
\begin{center}
\begin{minipage}[t]{16.5 cm}
\center
\epsfig{file=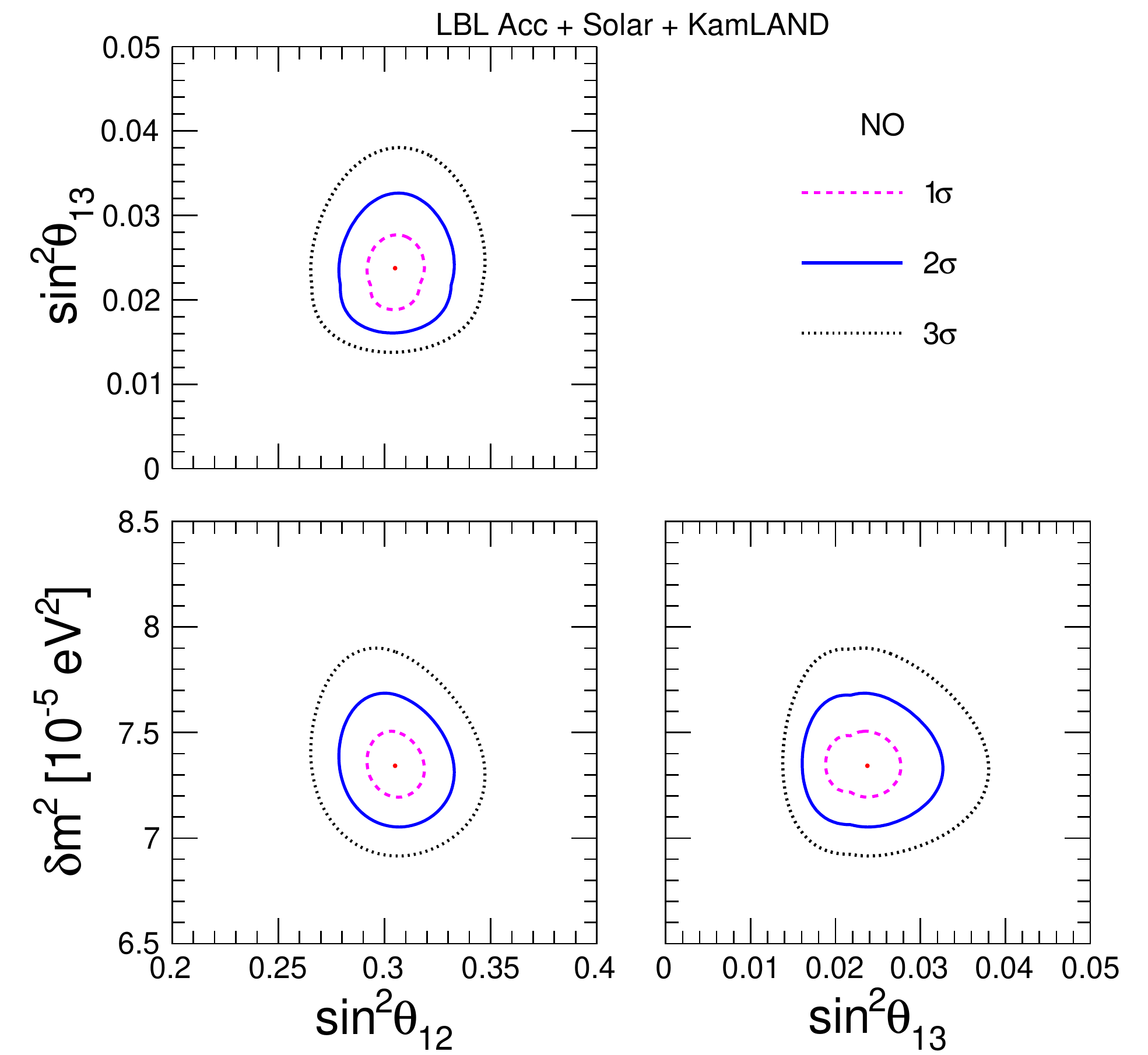,scale=0.35}
\end{minipage}
\begin{minipage}[t]{16.5 cm}
\caption{As in Fig.~5, but adding long-baseline accelerator data in the analysis.    
\label{fig_06}}
\end{minipage}
\end{center}
\end{figure}

\newpage

\begin{figure}[t]
\begin{center}
\begin{minipage}[t]{16.5 cm}
\center
\epsfig{file=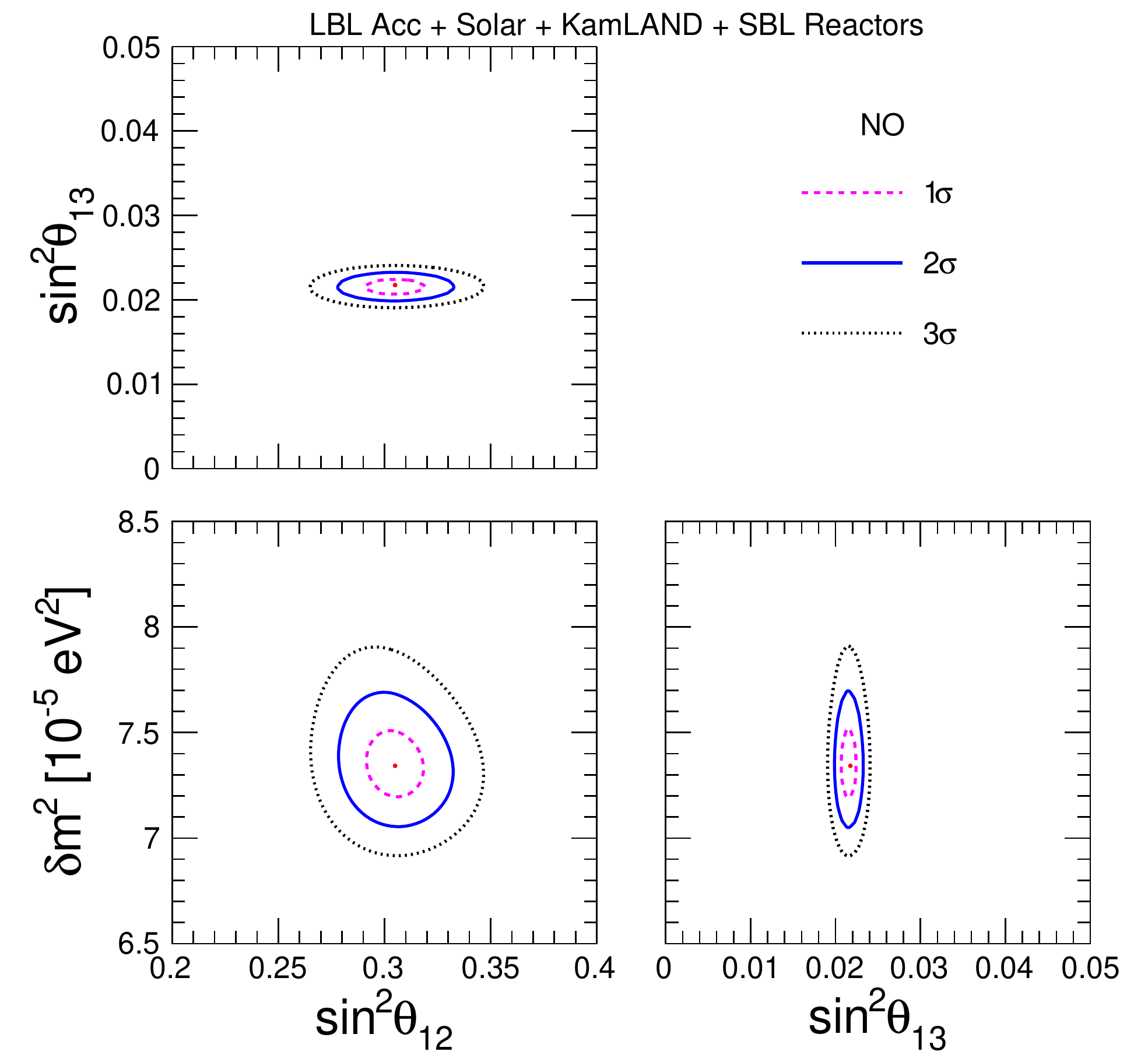,scale=0.35}
\end{minipage}
\begin{minipage}[t]{16.5 cm}
\caption{As in Fig.~6, but adding short-baseline rector data in the analysis.    
\label{fig_07}}
\end{minipage}
\end{center}
\end{figure}

Figure~7 is analogous to Fig.~6 but includes short-baseline reactor data, 
that provide further dramatic reduction of the allowed range for $\theta_{13}$
as compared with Figs.~5 and 6.
In this context, the inclusion of atmospheric neutrino data (not shown) would not 
induce any appreciable variation with respect to Fig.~7. A further reduction 
of the $\theta_{13}$ allowed range is expected from the final (and possibly combined) datasets collected by 
the running reactor experiments \cite{Meeting1,Meeting2}, while a very significant reduction
of the $(\delta m^2,\,\theta_{12})$ range will be possible in the medium-baseline 
JUNO experiment (in construction) \cite{An:2015jdp}.

\subsection{Covariances of $(\Delta m^2,\,\theta_{23},\theta_{13})$}

The covariances of parameter pairs among $(\Delta m^2,\,\theta_{23},\theta_{13})$  help to understand the 
interplay of different data sets in producing various single-parameter results discussed in  Section~3.
Since there are appreciable differences in NO and IO, we show both cases in
the following figures. 

Figure~8 shows the regions allowed at $N\sigma$ in the plane charted by $(\sin^2\theta_{23},\,\sin^2\theta_{13})$,
for both NO (upper panels) and IO (lower panels), for increasingly  rich data sets (panels from left to right).
The leading LBL appearance amplitude in Eq.~(\ref{Pacc}), governed by the $\sin^2\theta_{23}\sin^2\theta_{13}$,
induces an anti-correlation between these two parameters, visible in the left panels. Subleading effects
sensitive to $\mathrm{sign}(\Delta m^2)$ [second and third row of Eq.~(\ref{Pacc})] 
generate a difference in the allowed $\theta_{13}$ ranges for NO and IO,  the latter ones being generally higher. 
The middle panels show the combination with SBL reactor data.
The comparison of the left and middle panels shows that current accelerator and reactor constraints on $\theta_{13}$  
are more consistent in NO than in IO. This fact provides the  increment  of the $\Delta\chi^2$ difference
from Eq.~(\ref{chi1}) to Eq.~(\ref{chi2}). 
Atmospheric neutrino data cannot improve $\theta_{13}$ further,
but remain sensitive to the mass ordering and provide an independent  $\Delta\chi^2$ increment from 
Eq.~(\ref{chi2}) to Eq.~(\ref{chi3}). Concerning $\theta_{23}$, 
note that the slight $\theta_{13}$ tension between accelerator and reactor
data in IO; such tension is minimized for relatively large  $\theta_{23}$, hence the more pronounced preference for the second octant.
In any case, the octant ambiguity remains unresolved at $2\sigma$ level in both NO and IO.

\newpage

\begin{figure}[t]
\begin{center}
\begin{minipage}[t]{16.5 cm}
\center
\epsfig{file=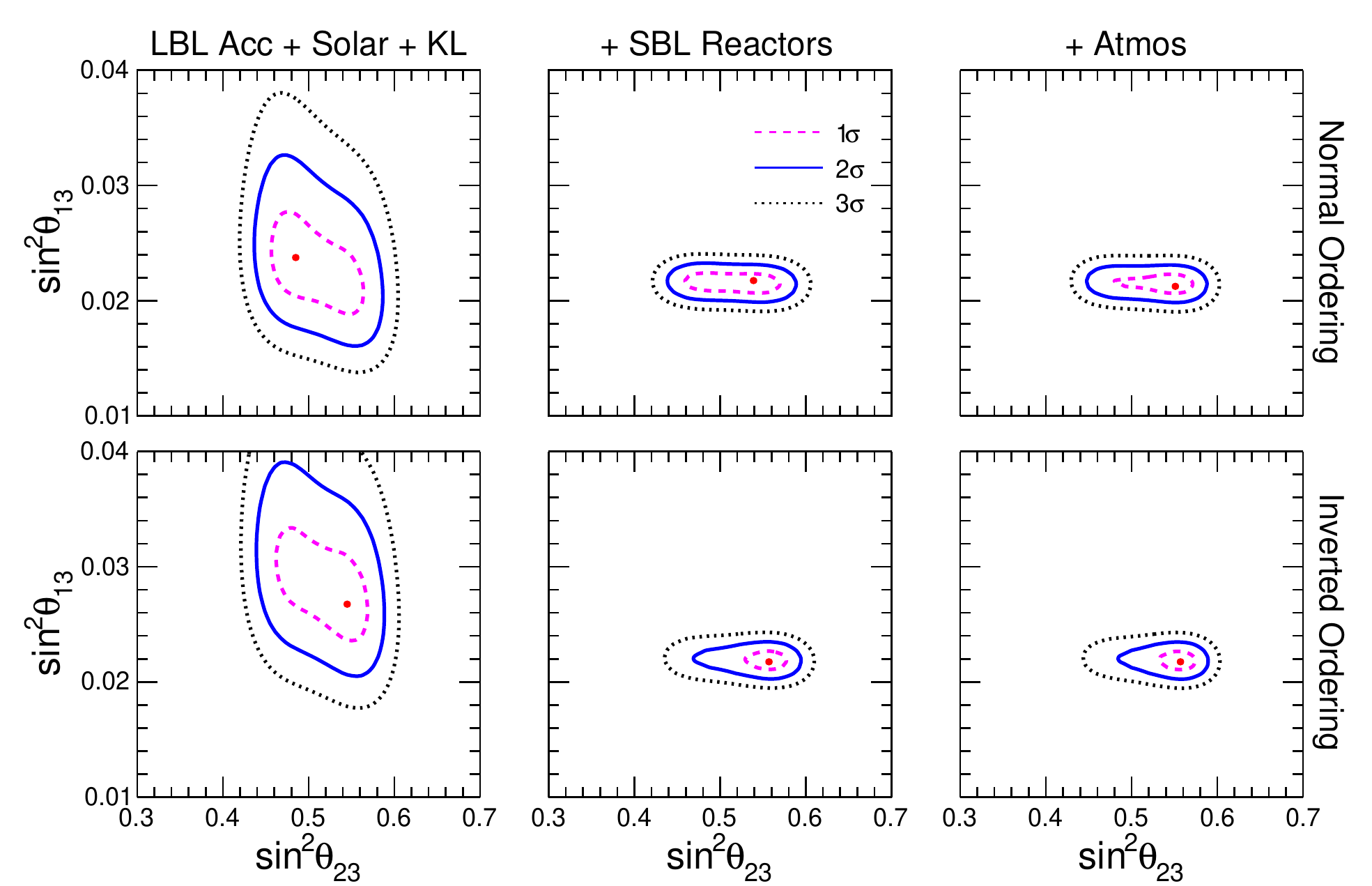,scale=0.7}
\end{minipage}
\begin{minipage}[t]{16.5 cm}
%\vspace*{-3mm}
\caption{Covariance plot for the $(\sin^2\theta_{13},\,\sin^2\theta_{23})$ parameters in NO (upper panels)
and IO (lower panels), as derived from an analysis of
LBL accelerator + solar + KL data (left panels), plus SBL reactor data (middle panels), plus 
atmospheric neutrino data (right panels).    
\label{fig_08}}
\end{minipage}
\end{center}
\vspace*{-20mm}
\end{figure}

\phantom{...}

\phantom{...}

\phantom{...}

\phantom{...}

\phantom{...}

\begin{figure}[h]
\begin{center}
\begin{minipage}[t]{16.5 cm}
\center
\epsfig{file=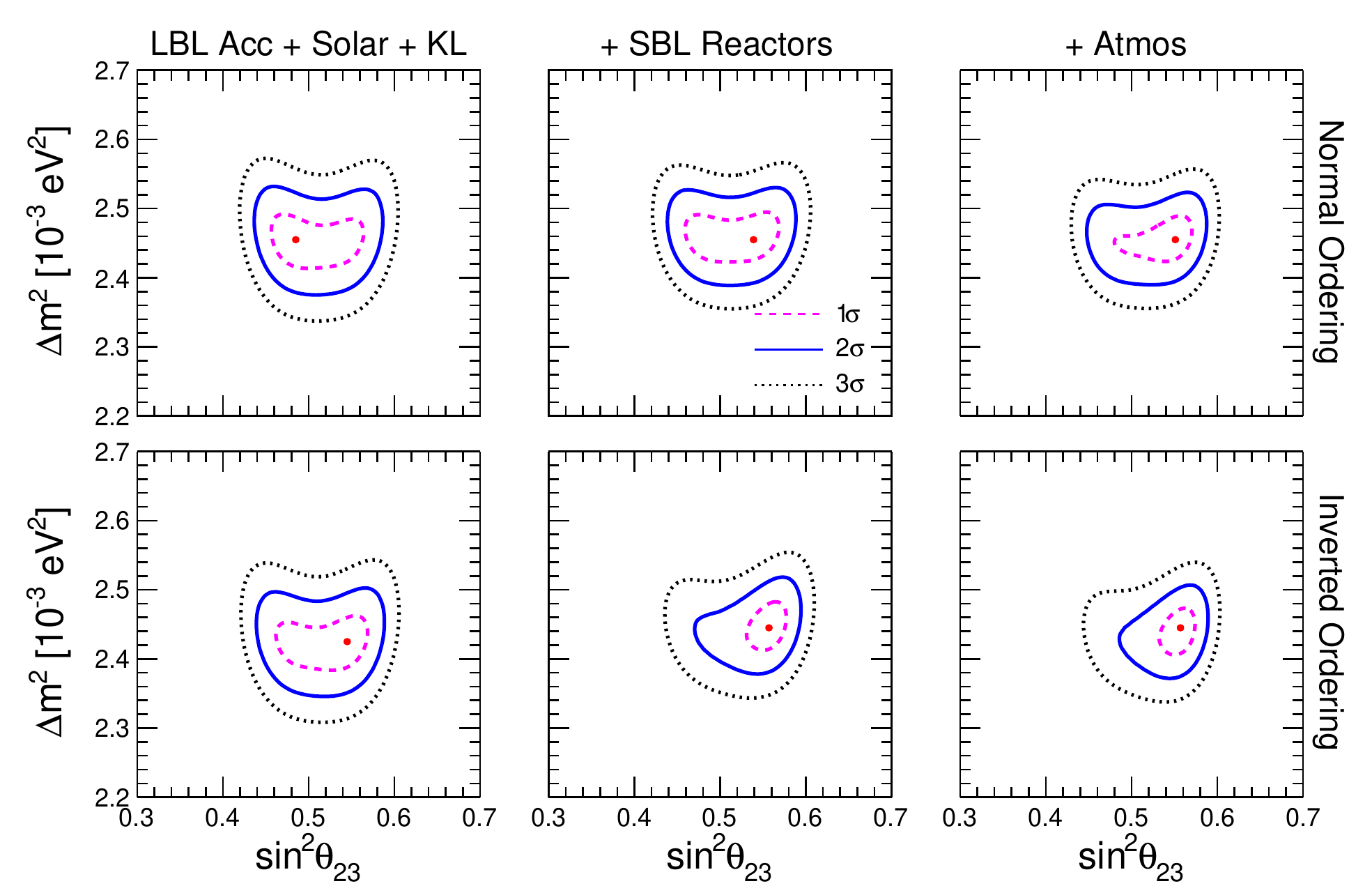,scale=0.7}
\end{minipage}
\begin{minipage}[t]{16.5 cm}
%\vspace*{-3mm}
\caption{Covariance of the $(\Delta m^2,\,\sin^2\theta_{23})$ parameters. 
\label{fig_09}}
\end{minipage}
\end{center}
\vspace*{-20mm}
\end{figure}

\newpage

\begin{figure}[t]
\begin{center}
\begin{minipage}[t]{16.5 cm}
\center
\epsfig{file=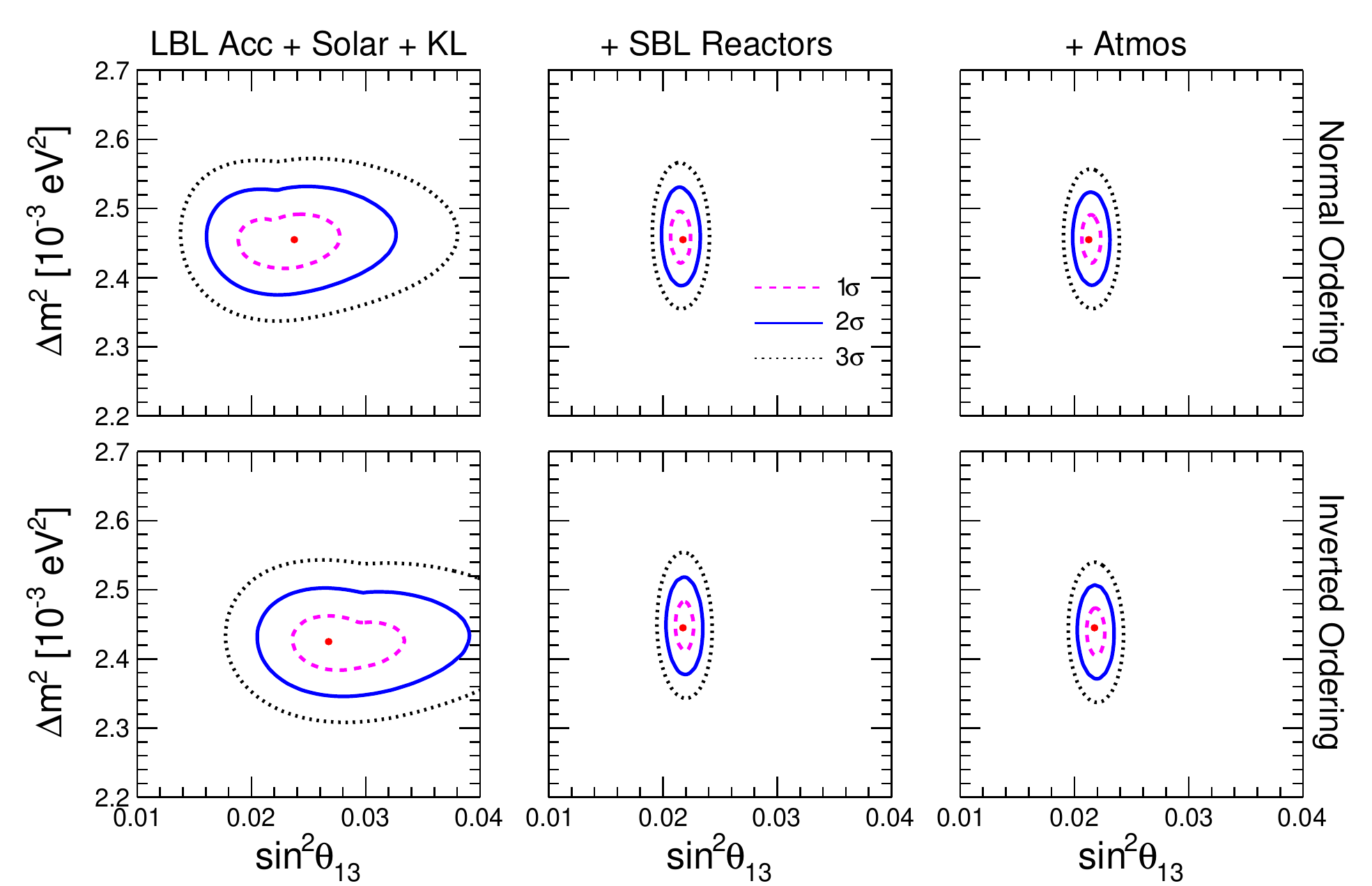,scale=0.7}
\end{minipage}
\begin{minipage}[t]{16.5 cm}
\vspace*{-3mm}
\caption{Covariance of the $(\Delta m^2,\,\sin^2\theta_{13})$ parameters. 
\label{fig_10}}
\end{minipage}
\end{center}
\vspace*{-10mm}
\end{figure}

Figure~9 shows the covariance of the $(\Delta m^2,\,\sin^2\theta_{23})$ parameters, in the same format as Fig.~8.
A striking feature of the NO case (upper panels)
is the very good consistency of all the data on $\Delta m^2$, whose best-fit value remains practically 
constant in the three upper panels. In IO (lower panels) the value of $\Delta m^2$ slightly increases
after the addition of SBL reactor data, as consequence of the slight increment in $\sin^2\theta_{23}$
discussed for Fig.~8. This small increase of $\Delta m^2$ slightly worsen the agreement with 
IC-DC data, that tend to prefer relatively low values of $\Delta m^2$ \cite{Aartsen:2017nmd}. 
Thus the IC-DC data set also provides a small contribution to the $\Delta \chi^2$ (about one unit) favoring
NO in Eq.~(\ref{chi3}).
Note that, in general, at nearly maximal mixing one gets the lowest allowed 
values of $\Delta m^2$, while for nonmaximal mixing (in either octants) the preferred values of 
$\Delta m^2$ tend to increase. This correlation stems mainly from disappearance data in LBL
accelerator experiments, where a decrease of the leading oscillation amplitude (governed by 
$\sin^22\theta_{23}$) can be partly traded for an increase of the leading oscillations phase (governed 
by $\Delta m^2$), so as to keep the disappearance rate nearly constant. 

Figure~10 shows the covariance of the $(\Delta m^2,\,\sin^2\theta_{13})$ parameters.
In the left panels, one can notice the mentioned preference for higher values of $\theta_{13}$ in IO, as 
well as a small kink in the contours. This feature shares the same origin as the kink discussed in the context of Fig.~6, 
namely, the contours correspond {to} two slightly displaced allowed ranges for $\theta_{13}$,
related to nearly-degenerate fits in the $\theta_{23}$ octants. Such kink
disappears {with the addition of} SBL reactor data.

Summarizing, Figs.~8--10 show the interplay among the $(\Delta m^2,\,\theta_{23},\theta_{13})$ parameters
within different data sets. In the NO case there is very good agreement among the values of both
$\Delta m^2$ and $\theta_{13}$ {from} different data {sets}, and $\theta_{23}$ remains nearly maximal at $1\sigma$,
with a minor preference for the second octant. In IO there is a slight tension between 
accelerator and reactor constraints on $\theta_{13}$, which contributes to disfavor IO and 
to slightly prefer the second octant of $\theta_{23}$ --- these two trends  being corroborated
by atmospheric data. However, even in IO, the octant degeneracy remains unresolved at $2\sigma$.

A final remark is in order. The previous
 description of fine details 
(at the $1\sigma$--$2\sigma$ level) in the covariances 
of the known parameters $(\Delta m^2,\,\theta_{23},\theta_{13})$ 
is based on the assumption of standard $3\nu$ oscillations. At the same level 
of detail, possible new neutrino physics (such as nonstandard interactions discussed before)
 might induce noticeable changes and  could
shift the best fits, alter the various hints, or spoil apparent data agreement. 
The fragility of fit details (at $1\sigma$--$2\sigma$ level) under such possible perturbations 
 should always be kept in mind.

\newpage

\begin{figure}[t]
\begin{center}
\begin{minipage}[t]{16.5 cm}
\center
\epsfig{file=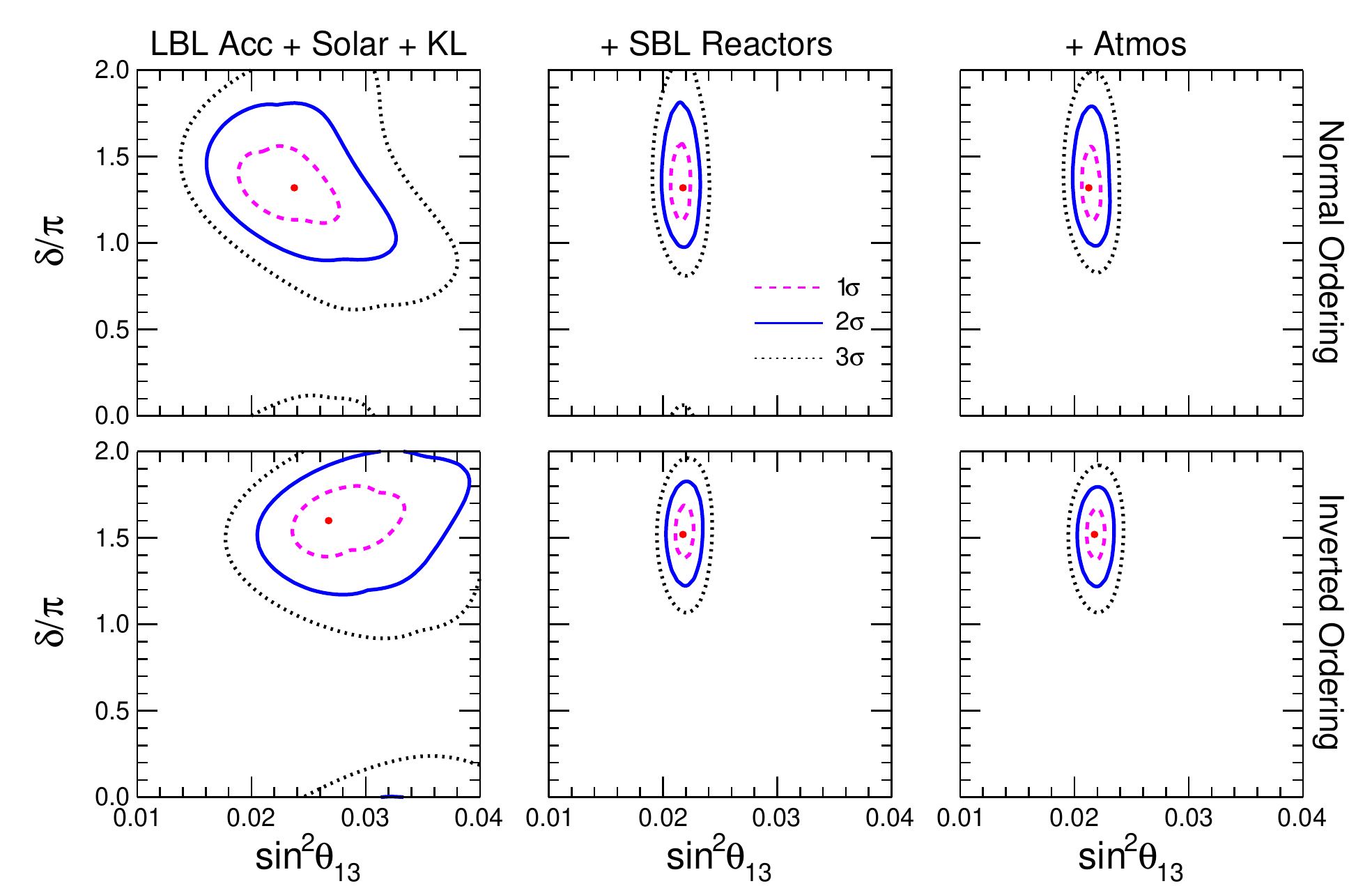,scale=0.7}
\end{minipage}
\begin{minipage}[t]{16.5 cm}
%\vspace*{-3mm}
\caption{Covariance of the $(\delta,\,\sin^2\theta_{13})$ parameters. 
\label{fig_11}}
\end{minipage}
\end{center}
\vspace*{-20mm}
\end{figure}

\phantom{...}

\phantom{...}

\phantom{...}

\phantom{...}

\phantom{...}

\begin{figure}[h]
\begin{center}
\begin{minipage}[t]{16.5 cm}
\center
\epsfig{file=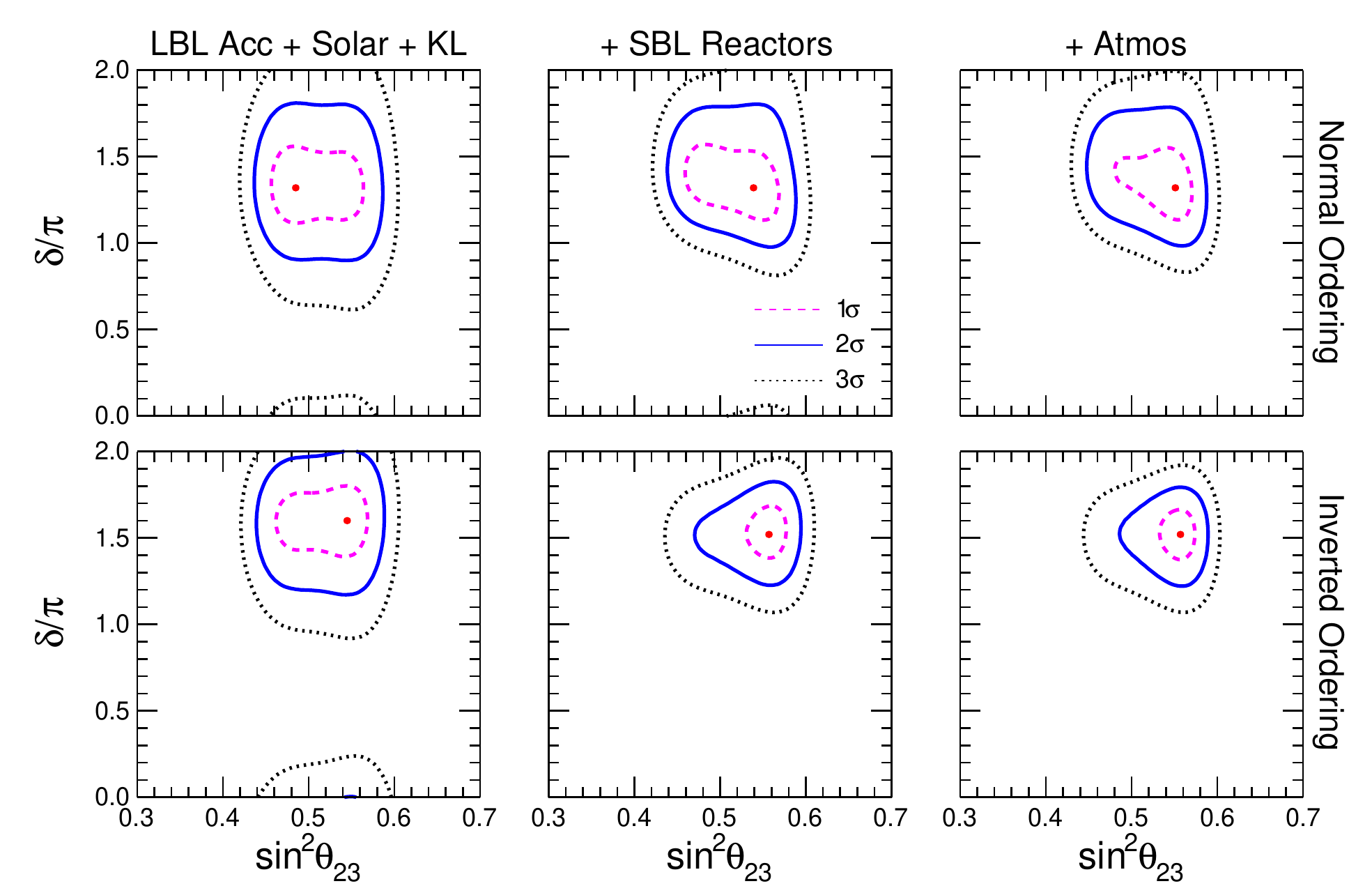,scale=0.7}
\end{minipage}
\begin{minipage}[t]{16.5 cm}
%\vspace*{-3mm}
\caption{Covariance of the $(\delta,\,\sin^2\theta_{23})$ parameters. 
\label{fig_12}}
\end{minipage}
\end{center}
\vspace*{-20mm}
\end{figure}

\newpage

\subsection{Covariances involving $\delta$}

We complete our analysis of covariances by discussing those involving the 
unknown parameter $\delta$ and one of the two known mixing angles $(\theta_{13},\,\theta_{23})$.
There is a vast literature on analytical studies of such correlations, the works in 
\cite{Coloma:2014kca,Ghosh:2015ena,Lindner:2017nmb} providing just a few recent examples among many. Here we focus on the 
phenomenological correlations stemming from current data.

Figure~11 shows the $N\sigma$ regions allowed in the plane 
$(\delta,\,\sin^2\theta_{13})$. The strong correlations between these two 
parameters (in the left panels)
are mainly induced by the interplay between $\delta$ and $\theta_{13}$ arising in 
the subleading terms (second and third row) of Eq.~(\ref{Pacc}).  In NO, 
the best fit of $\delta$ remains very close to $\sim 1.3\pi$ by adding first SBL reactor and 
then atmospheric neutrino data. In NO, the consistency of all the data sets towards the same 
best-fit values of both the known $(\Delta m^2,\,\theta_{23},\theta_{13})$ parameters 
and of the unknown $\delta$ phase is striking. In IO there is a slight 
decrease of $\delta$ from left to middle panels, correlated to the decrease 
of $\theta_{13}$. 

Figure~12 shows the $(\delta,\,\sin^2\theta_{23})$ covariance. Only weak correlations (if any) emerge 
between these two parameters at the current level of accuracy; see, e.g., \cite{Lindner:2017nmb} for prospective improvements. 
In particular, in NO there is a slight anti-correlation,
which implies that the best fit of $\delta$ might increase from $\sim 1.3$
to $\sim 1.4$ if the first (rather than the second) octant of $\theta_{23}$ 
were favored by upcoming data. In IO there is no significant correlation.
These considerations about the interplay among three unknowns (the phase $\delta$, the $\theta_{23}$ octant and 
the mass ordering) are rather fragile and might change with future data. Conversely, the
overall $3\sigma$ constraints on $\delta$ emerging in Fig.~11 and 12 appear to
be relatively robust,  with modest dependence on  $(\theta_{13},\,\theta_{23})$. 

In a sense, the CP phase is already being ``measured''
by current experiments, with an effective $1\sigma$ accuracy of $\sim 15\%$ in NO and $\sim 9\%$ in IO 
(see also Table~1). In the favored case of NO, this accuracy is sufficient to reject
the CP-conserving case $\delta=0$ (or $2\pi)$ at $3\sigma$, but is not enough to exclude the
other CP-conserving case  $\delta=\pi$ at $2\sigma$.  Both cases are instead  excluded at $3\sigma$ in the IO case that, however,
is in turn disfavored with respect to NO case, see Eq.~(\ref{chi3}). Summarizing,
although CP conservation cannot be rejected yet with significant confidence,
relatively stringent constraints on $\delta$ can be obtained from current data, with a clear preference 
for $\sin\delta <0$.

The progress made by
CP-sensitive oscillation searches is impressive:
in contrast, a dozen years ago  the CP-conserving cases were found 
to be largely degenerate at $<1\sigma$ level, with no significant difference between NO and IO \cite{Fogli:2005cq}. 
If this exciting trend continues, there are good prospects to eventually assess
CP violation at $>3\sigma$ with future data.

We conclude this section with some comments on scenarios beyond $3\nu$. 
As we have seen, there is a high degree of convergence of all the data within the 
standard $3\nu$ framework, which implies that new physics effects, if any, must remain
small as compared with the estimated uncertainties. In particular,  apart from
the mild tension between solar and KL results discussed earlier, there are no
relevant anomalous results pointing towards new neutrino interactions, 
whose couplings and other features can be thus constrained by oscillation (plus other) data. 

The situation is somewhat different concerning new neutrino states, since there
are long-standing anomalies that seem to point towards a 4th  massive (so-called sterile)
 state at the $O(1)$~eV scale, 
slightly mixed with both $\nu_e$ and $\nu_\mu$; see \cite{Gariazzo:2015rra,Abazajian:2012ys} for reviews.
This exciting possibility
is balanced by internal tensions between different data sets (especially appearance versus disappearance)
within the $4\nu$ oscillation scenario and its variants;
see \cite{Gariazzo:2017fdh,Dentler:2018sju} for recent goodness-of-fit assessment and parameter estimates. 
A large experimental investment is being made to clarify this puzzling situation, and 
interesting new results are expected  by a variety of short-baseline oscillation searches \cite{Spitz:2015gga,Lhuillier:2015fga}.
In this context, it is worth reminding that the $3\times 3$ matrix in Eq.~(\ref{PMNS}) 
would become non-unitary in the presence of new states at the eV scale or higher,
with new mixing angles and CP phases inducing effects degenerate with the standard $3\nu$ 
ones, see e.g.\ \cite{Barry:2011wb,Antusch:2006vwa,Parke:2015goa,Tang:2017khg,Li:2018jgd,Blennow:2016jkn,Fong:2017gke,Li:2015oal}. 
In particular, the determination of $\delta$, the discrimination of the mass hierarchy
and the resolution of the octant of $\theta_{23}$, appear 
to be quite sensitive to such additional unknowns, see 
\cite{Miranda:2016wdr,Klop:2014ima,Palazzo:2015gja,Capozzi:2016vac,Agarwalla:2016xlg,Agarwalla:2016mrc,Dutta:2016eks} 
for recent studies. 
Settling the
status of sterile neutrino oscillations will thus be beneficial to gain more confidence 
in current hints about $3\nu$ unknowns.

%%%%%%%%%%%%%%%%%%%%%%%%%%%%%%%%%%%%%%%%%%%%%%%%%%%%%%%%%%%%%%%%%%%%%%%%%%%%%%%%%%%%%%%%%%%%%%%%%%%%%%%%%
\section{Constraints from non-oscillation data and combination with oscillation searches}

Nonoscillation data from single $\beta$ decay,  $0\nu\beta\beta$ decay and cosmology 
 are crucial to probe the nature and absolute  masses of neutrinos, which
are not accessible via flavor oscillations. They also offer additional 
handles to probe the neutrino mass ordering and to check the consistency
of $3\nu$ framework \cite{Fogli:2004as,Fogli:2006yq}. 
Within the $3\nu$ scenario, the available 
beta-decay bounds are the level $m_\beta < 2$~eV \cite{Petcov}, while  
typical mass bounds placed by  $0\nu\beta\beta$ decay and cosmology are 
the sub-eV level \cite{Vogel,Olive,Lesg}. Therefore
we shall consider only the experimental 
bounds on $\Sigma$ and on $m_{\beta\beta}$ (the latter being valid if neutrinos are Majorana).

We follow the same $3\nu$ (frequentist) methodology as in \cite{Capozzi:2017ipn}, based on the construction 
of $\chi^2$ functions for $m_{\beta}$ and $\Sigma$, to be added to the $\chi^2$
function coming from the previous oscillation data analysis, marginalized over
all the known and unknown mass-mixing parameters and phases. There are
alternative approaches to absolute mass observables 
based on Bayesian statistics (see, e.g., \cite{Gerbino:2016ehw,Gariazzo:2018pei,Caldwell:2017mqu,Agostini:2017jim}), 
whose results depend somewhat
on prior assumptions.  As already remarked for oscillation observables we underline that, irrespective of statistical
details, possible new neutrino states and interactions (not considered herein) might profoundly 
affect our understanding of non-oscillation observables \cite{Mohapatra:2005wg}.

\subsection{Inputs from neutrinoless double beta decay and cosmology}

Running $0\nu\beta\beta$ experiments have not found evidence for this rare process so far 
\cite{DellOro:2016tmg,KamLAND-Zen:2016pfg,Albert:2017owj,Alduino:2017ehq,Aalseth:2017btx,Agostini:2018tnm},
and the quest for the neutrino nature (Dirac or Majorana) remains open. 
If neutrinos are Majorana,
within the $3\nu$ framework one can translate lower limits on the decay half life into upper limits
on $m_{\beta\beta}$, via the knowledge of the nuclear matrix element (NME) for the considered
nucleus \cite{DellOro:2016tmg,Vergados:2016hso,Vogel} 
Improving the NME calculations and the underlying nuclear models is imperative
to get significant constraints on $m_{\beta\beta}$ \cite{Engel:2016xgb}. In this context, the poorly known 
value of the effective axial coupling $g_A$ in the nuclear medium is being increasingly recognized as one of the most
serious issues in the field \cite{DellOro:2016tmg,Engel:2016xgb,Iachello:2015ejm},  
to be addressed with a variety of theoretical and experimental
tools also involving other weak-interaction nuclear processes \cite{Suhonen:2017krv}. 

Despite the relatively large uncertainties on the NME's,  
it is fair to say that the leading upper limit on $m_{\beta\beta}$ is currently 
placed by the KamLAND-Zen experiment using $^{136}$Xe \cite{KamLAND-Zen:2016pfg}. We convert its data into 
a function $\chi^2(m_{\beta\beta})$ according to the procedure in \cite{Capozzi:2017ipn}, which
marginalizes away the
conservative NME errors (including $g_A$ uncertainties) estimated in \cite{Lisi:2015yma}.
As a representative result we quote the $2\sigma$ upper limit
(which applies to both NO and IO) \cite{Capozzi:2017ipn},
%--------
\begin{equation}
m_{\beta\beta} < 0.18 \ \mathrm{eV\ at \ }2\sigma \ .
\end{equation}
%--------

Concerning cosmology, we adopt the same experimental inputs and analysis results reported in 
the recent paper \cite{Capozzi:2017ipn}. The analysis was based on six combinations of data coming from the so-called
TT, TE and EE anisotropy angular power spectra of Planck \cite{Aghanim:2015xee}, where T and E refer to temperature 
and polarization, respectively. Such Planck data were eventually supplemented with  
lensing potential power spectrum reconstruction data, and with
optical depth HFI constraints ($\tau_\mathrm{HFI}$) \cite{Aghanim:2016yuo}. 
Also considered were baryon acoustic  
oscillation (BAO) measurements \cite{Beutler:2011hx,Ross:2014qpa,Anderson:2013zyy}. 
The adopted cosmological framework was based on the so-called $\Lambda$CDM model, with allowance
for massive neutrinos ($\Lambda$CDM+$\Sigma$).
Systematic uncertainties affecting the $\Lambda$CDM+$\Sigma$ model
were lumped in a dominant parameter $A_\mathrm{lens}$ (with significant covariance with $\Sigma$),
that was optionally left free to vary around its standard value 
($A_\mathrm{lens}=1$), in  order to improve the overall fit of Planck lensing data \cite{Ade:2015zua}.

In the work \cite{Capozzi:2017ipn}, a total of 6+6 data sets (with and without free $A_\mathrm{lens}$)
were thus considered. In combination with oscillation data, 
upper limits at $2\sigma$ were obtained for these 12 cases, 
mostly in the sub-eV range for the $\Lambda$CDM+$\Sigma$ model, with 
somewhat weaker results for $\Lambda$CDM+$\Sigma$+$A_\mathrm{lens}$ variant. 
Allowance was given  for different (non-degenerate) neutrinos masses, 
inducing small differences between the
overall $\chi^2$ in NO and IO at small $\Sigma$.
Interestingly,  NO was generally favored over IO, 
although only by a fraction of $\Delta\chi^2$ unit in typical cases. 
These results (in particular, the numerical values in  Table~II of \cite{Capozzi:2017ipn}) 
are confirmed by including the updated oscillation data discussed
above, and are not repeated. 

Among the twelve cases reported in \cite{Capozzi:2017ipn}, we discuss here only the two cases 
labelled 1 and 6 in the $\Lambda$CDM+$\Sigma$ model. They 
provide representative example of ``weak'' cosmological upper limits 
(just below the eV scale), and ``strong'' cosmological upper limits (in the
sub-eV range). The corresponding
$\chi^2$ functions  are taken from \cite{Capozzi:2017ipn} for both NO and IO,
and provide the following $2\sigma$ upper limits: 
%----------------
\begin{eqnarray}
\mathrm{``weak"\ limit} &:& \Sigma<0.72\ \mathrm{(NO)}\ \mathrm{or\ } \Sigma<0.80\ \mathrm{(IO)}\  \mathrm{at\ }2\sigma\ ,\\
\mathrm{``strong"\ limit} &:& \Sigma<0.18\ \mathrm{(NO)}\ \mathrm{or\ } \Sigma<0.20\ \mathrm{(IO)}\  \mathrm{at\ }2\sigma\ .
\end{eqnarray}
%----------------
As already emphasized, we discuss NO and IO separately,  
and do not consider anymore the marginalized ``any ordering'' case \cite{Capozzi:2017ipn},
which would display only NO regions (and no IO region) up to $3\sigma$.

\begin{figure}[t]
\begin{center}
\begin{minipage}[t]{16.5 cm}
\center
\epsfig{file=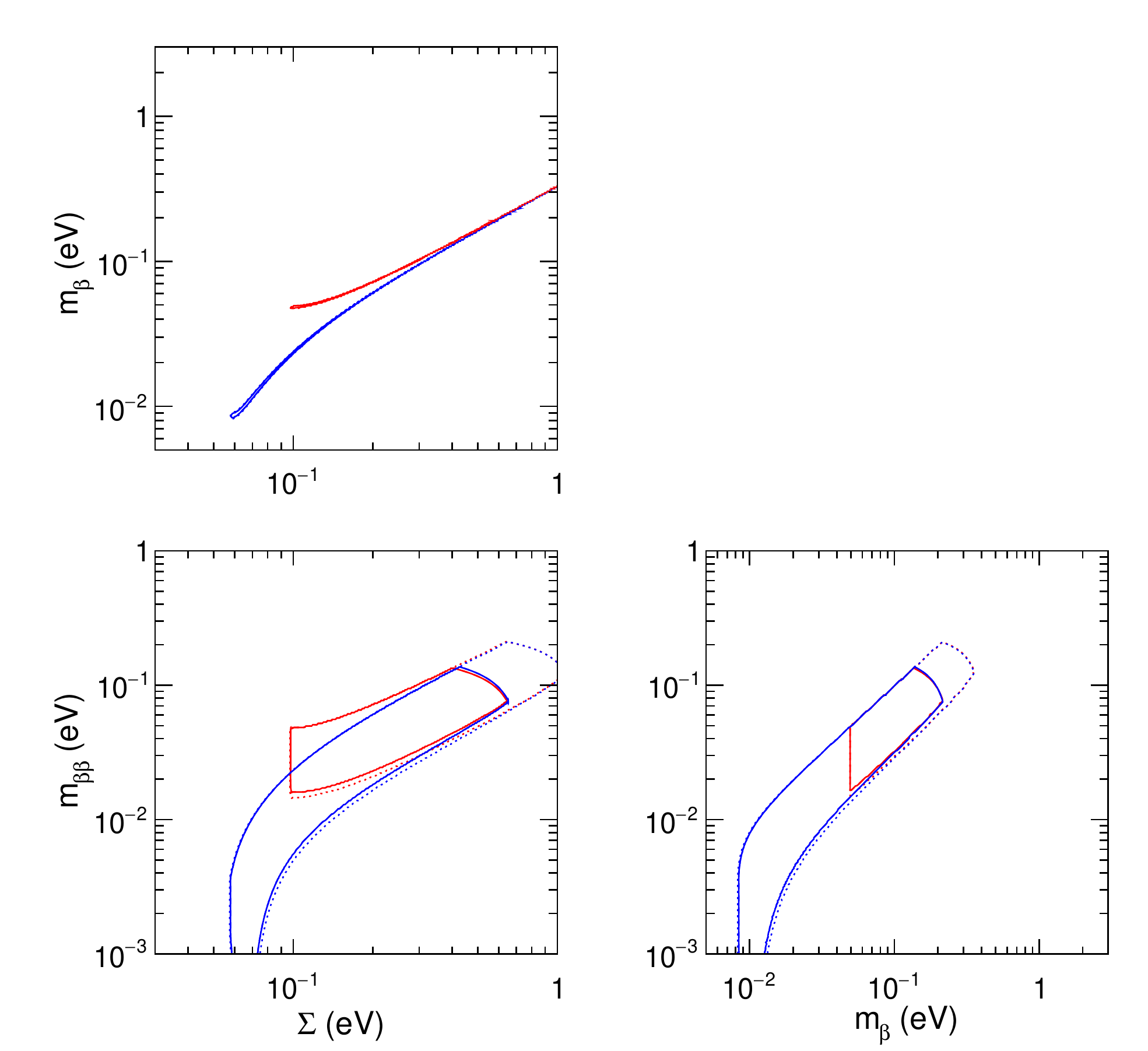,scale=0.56}
\end{minipage}
\begin{minipage}[t]{16.5 cm}
\vspace*{-3mm}
\caption{Combined $3\nu$ analysis of oscillation and nonoscillation data, in the planes charted by any pair among 
the absolute mass observables $(m_\beta,\,m_{\beta\beta},\,\Sigma)$. Bounds from $0\nu\beta\beta$ are derived
from KamLAND-Zen data and NME estimates.  Bounds from cosmology refer to the
representative ``weak'' limit  described in  the text. The allowed bands correspond to $N\sigma=2$ (solid) and 
$N\sigma=3$ (dotted), for both NO (blue) and IO (red), taken as separate cases. 
If the $\Delta\chi^2_\mathrm{IO-NO}$ difference in Eq.~(\protect\ref{chi3})
were included, the IO bands would disappear.} 
\label{fig_13}
\end{minipage}
\end{center}
\vspace*{-5mm}
\end{figure}

\subsection{Representative bounds on $(m_\beta,\,m_{\beta\beta},\,\Sigma)$}

Figure~13 shows the results of a combined $3\nu$ analysis 
of oscillation and nonoscillation data, in the planes charted by any pair among 
the absolute mass observables $(m_\beta,\,m_{\beta\beta},\,\Sigma)$, for the
``weak'' cosmological limit described above. The allowed bands correspond to $2\sigma$ (solid) and 
$3\sigma$ (dotted), for both NO (blue) and IO (red). Note the spread in $m_{\beta\beta}$ at any fixed
value of $m_\beta$ or of $\Sigma$, induced by the unknown Majorana phases.  
In the plane $(\Sigma,\,m_{\beta\beta})$, there
is an interesting sinergy between the upper bounds on these two parameters, coming from cosmological
and $0\nu\beta\beta$ data, respectively: the first would cut the allowed band vertically, while
the second horizontally, their combination providing a ``slanted'' upper limit to the allowed band.
The other two panels show also the projection onto the $m_\beta$ variable, with $2\sigma$ upper limits
slightly above $0.2$~eV. This fraction of the $m_\beta$ allowed range can be probed by the
KATRIN experiment (in construction) \cite{Katrin}.

\begin{figure}[t]
\begin{center}
\begin{minipage}[t]{16.5 cm}
\center
\epsfig{file=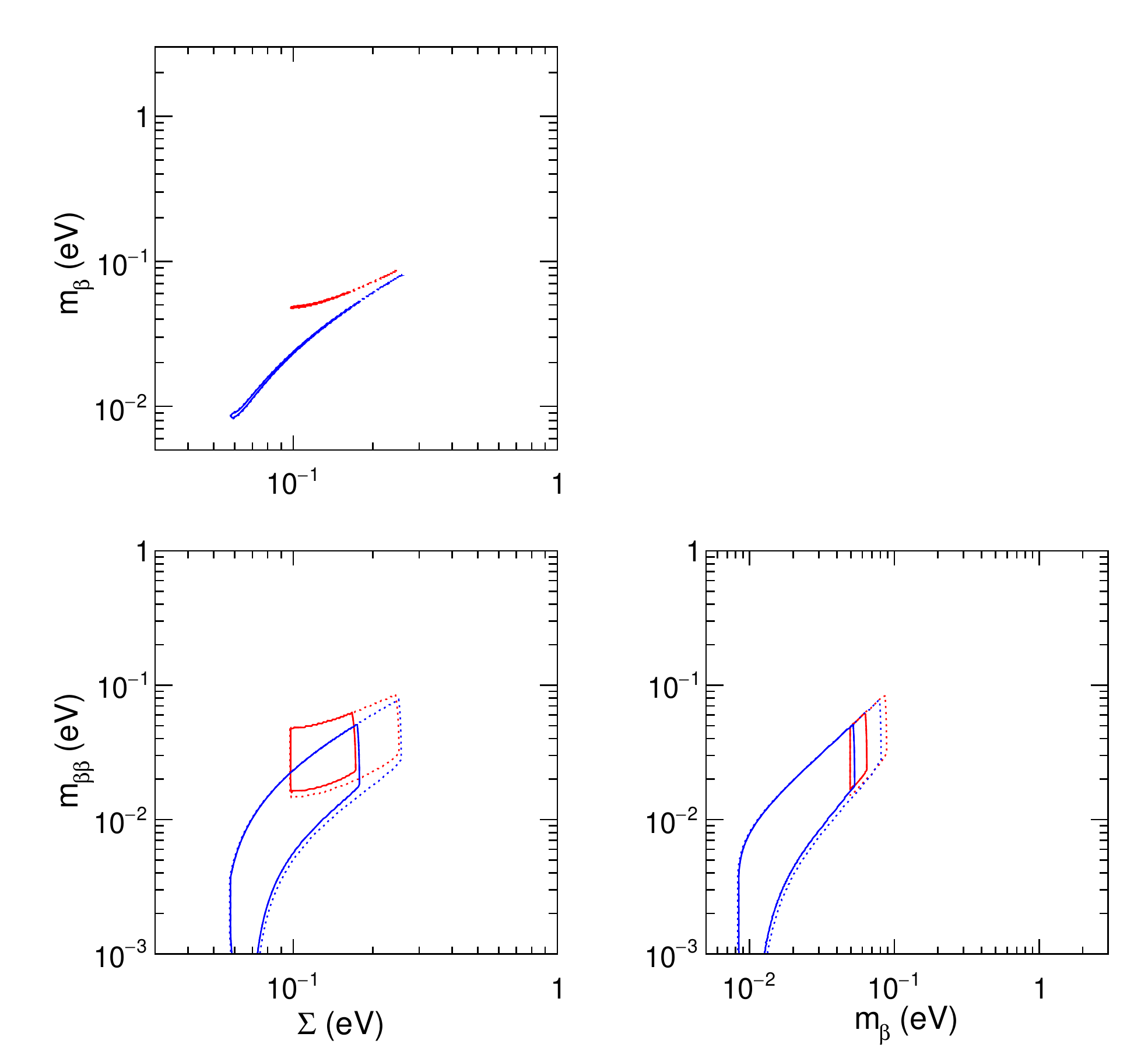,scale=0.56}
\end{minipage}
\begin{minipage}[t]{16.5 cm}
\vspace*{-3mm}
\caption{As in Fig.~13, but for ``strong'' cosmological limits as
described in  the text.} 
\label{fig_14}
\end{minipage}
\end{center}
\vspace*{-4mm}
\end{figure}

Figure~14 is analogous to Fig.~13, but  refers to the
``strong'' cosmological limit described above. This limit dominates over the
$0\nu\beta\beta$ bound in the fit and, in the $(\Sigma,\,m_{\beta\beta})$ plane,
it provides a vertical cut to the allowed bands. In the 
$(\Sigma,\,m_{\beta})$ plane, the narrow bands for NO and IO are 
completely separated. At least in principle,  precise 
$(\Sigma,\,m_{\beta})$ measurements could then be able to select 
one mass ordering. 
Unfortunately, the allowed $m_\beta$ range
is well below the KATRIN sensitivity, although it might be partly
accessed with future projects based on new detector concepts 
\cite{Robertson:2015owa,Mertens:2016ihw,Gastaldo:2017hgn}. Note that
cosmology could select NO at $2\sigma$, if the upper bound were reduced by
a factor of two.

Summarizing, measurements of $(m_\beta,\,m_{\beta\beta},\,\Sigma)$ have 
the potential to test the $3\nu$ paradigm and its three mass-related
unknowns: the fundamental nature of the mass term, 
the absolute neutrino mass scale, and the mass ordering.
The first unknown remains as such, both options (Dirac or Majorana) being possible
in the absence of a $0\nu\beta\beta$ decay signal.
The second remains also undetermined, but with upper limits which are steadily
decreasing and will eventually hit the signal. The third unknown is being
approached by cosmology, although only weakly at present---the preference for NO
being mainly driven by current oscillation data.  
 \newpage

%%%%%%%%%%%%%%%%%%%%%%%%%%%%%%%%%%%%%%%%%%%%%%%%%%%%%%%%%%%%%%%%%%%%%%%%%%%%%%%%%%%%%%%%%%%%%%%%%%%%%%%%%
\section{Summary and conclusions}

We have presented 
an up-to-date global analysis of data coming from neutrino oscillation and non-oscillation experiments, 
as available in April 2018, within the standard framework including three massive and mixed neutrinos. We 
have discussed in
detail the status of the three-neutrino ($3\nu)$ mass-mixing parameters, both known and unknown,
as listed in Eqs.~(\ref{eq:knowns}) and (\ref{eq:unknowns}). 

The main results from the analysis of oscillation searches are summarized graphically in Fig.~3 and
numerically in Eq.~(\ref{chi3}) and Table~1. Concerning the known parameters:
the squared mass differences $(\delta m^2,\,|\Delta m^2|)$ 
are determined within a couple of percent, while the mixing parameters 
$(\sin^2\theta_{12},\,\sin^2\theta_{13},\,\sin^2\theta_{23})$
within a few percent, see the last column in Table~1 for more precise values.
Concerning the unknown parameters: a preference for NO emerges at $3\sigma$ level from the global
analysis, with coherent contributions from various data sets. If the $\Delta \chi^2$ difference in Eq.~(\ref{chi3}) is taken at face value, no allowed region survives for IO up to $3\sigma$.
By considering NO and IO as separate cases, we also find that 
the Dirac CP phase $\delta$ is constrained within $\sim 15\%$ ($\sim 9\%$) uncertainty in NO (IO) around nearly-maximal CP-violating values, $\delta\sim 3\pi/2$. The CP-conserving value $\delta =0$ (or $2\pi)$ is
disfavored at $3\sigma$ in both NO and IO; the  value $\delta=\pi$ is also disfavored at $3\sigma$ in IO
but not in NO (where it is still allowed at $2\sigma$). Concerning deviations of $\theta_{23}$ from 
maximal mixing, we find an overall preference for the second octant (more pronounced in IO), although
both octants are allowed at $2\sigma$.  

The above results have been discussed in detail in terms of increasingly rich data sets and of
 covariance plots between various pairs
of parameters. We have also tried to convey the message that oscillation data analyses are becoming increasingly
complicated to be performed outside the experimental collaborations. External users 
may need to adopt officially
processed results, e.g., in terms of $\chi^2$ maps when available. In this work, we have used such maps 
for Day Bay reactor results, as well as for Super-Kamiokande and IceCube-DeepCore atmospheric results. 
However, the integration of raw data and processed results should always be performed in a critical way. 
We have argued that progress in the field of data analyses requires an advanced  
discussion of theoretical and experimental uncertainties, at the same level
of refinement of other mature fields in particle physics. Such progress is crucial
to probe further the emerging hints  on the unknown $3\nu$ oscillations parameters. We have also
remarked that these hints may be perturbed by  possible new states and interactions 
beyond the standard $3\nu$ framework.

Concerning the  non-oscillation observables 
$(m_\beta,\,m_{\beta\beta},\,\Sigma)$, the combination with oscillation constraints has been shown for two
representative cases, corresponding to ``weak'' and ``strong'' bounds from cosmology, in Figs.~13 and 14 respectively. In the absence of a $0\nu\beta\beta$ signal, the Dirac or Majorana nature of massive
neutrinos remains undetermined. 
The absolute neutrino mass scale is also undetermined, but with upper limits which are steadily
decreasing. For ``weak'' cosmological bounds,
a fraction of the allowed regions can be probed by direct mass searches. Limits from cosmology appear rather promising, as a reduction of the  ``strong'' bounds by
a factor of two might approach the threshold for NO-IO separation. 

In conclusion, the $3\nu$ mass-mixing paradigm is being probed with increasing accuracy. The known oscillation
parameters are determined with a few percent precision, and statistically significant hints are emerging
on the mass ordering and on the CP-violating Dirac phase. The progress made in a dozen years
since a previous review in this Journal \cite{Fogli:2005cq} is impressive. We look forward to seeing
the completion of the $3\nu$ framework, as well as hints (and possible signals) coming from absolute mass observables in the future. 
Possible indications of new physics beyond the $3\nu$ paradigm might also emerge, and provide surprising and novel
directions for global analyses.

\section*{Acknowledgments}

We thank E.\ Di Valentino and A.\ Melchiorri for permission to use  the cosmological 
likelihood functions previously 
derived and presented in \cite{Capozzi:2017ipn}. We are grateful to various members of experimental collaborations for informing us about
the public release of their official neutrino oscillation data analyses, including: 
M.~Nakahata, T.~Nakaya and D.~Wark (Super-Kamiokande atmospheric data),
E.~Resconi and T.~DeYoung (IceCube DeepCore atmospheric data), and 
D.~Naumov (Daya Bay reactor data). 

A.P. is supported by the grant ``Future In Research'' {\em Beyond three neutrino families}, 
Fondo di Sviluppo e Coesione 2007-2013, APQ Ricerca Regione Puglia, Italy, 
``Programma regionale a sostegno della specializzazione intelligente e della sostenibilit\`a sociale ed ambientale''. 
E.L., A.M. and A.P. acknowledge partial support by the research project TAsP 
(Theoretical Astroparticle Physics) funded by the Instituto Nazionale di Fisica Nucleare (INFN).
F.C.\ acknowledges partial support by the Deutsche Forschungsgemeinschaft
through Grant No.\ EXC 153 (Excellence Cluster ``Universe'') and Grant No.\
SFB 1258 (Collaborative Research Center ``Neutrinos, Dark Matter,
Messengers'') as well as by the European Union through Grant
No.\ H2020-MSCA-ITN-2015/674896 (Innovative Training Network ``Elusives'').

\end{document}